\documentclass[fleqn,usenatbib]{mnras}

\usepackage{newtxtext,newtxmath}

\usepackage[T1]{fontenc}
\usepackage[utf8]{inputenc}
\DeclareUnicodeCharacter{2212}{-}

\usepackage[table,xcdraw]{xcolor}
\usepackage{ulem}
\usepackage{float}

\DeclareRobustCommand{\VAN}[3]{#2}
\let\VANthebibliography\thebibliography
\def\thebibliography{\DeclareRobustCommand{\VAN}[3]{##3}\VANthebibliography}

\def\alstar{{\textsc a}l{\textsc s}tar}    

\newcommand{\fwhm}{{\sc fwhm}}
\newcommand{\class}{{\sc class\_star}}
\newcommand{\fluxradius}{{\sc flux\_radius}}
\newcommand{\spread}{{\sc spread\_model}}

\usepackage{xspace,color}

\usepackage{graphicx}	
\usepackage{amsmath}	

\title[S+FP: A first 12-band glimpse of Fornax]{The S-PLUS Fornax Project (S+FP): A first 12-band glimpse of the Fornax galaxy cluster}

\author[Smith Castelli et al.]{
A.V. Smith Castelli,$^{1,2}$\thanks{E-mail: asmith@fcaglp.unlp.edu.ar (AVSC)}
A. Cortesi,$^{3,4}$
R.F. Haack,$^{1,2}$
A.R. Lopes,$^{1}$
J. Thainá-Batista,$^{5}$
R. Cid Fernandes,$^{5}$
\newauthor
L. Lomel\'i-N\'u\~nez,$^{3}$
U. Ribeiro,$^{6}$
C.R. de Bom,$^{6}$ 
V. Cernic,$^{7}$  
L. Sodré Jr,$^{7}$
L. Zenocratti,$^{1,2}$
M.E. De Rossi,$^{8,9}$
\newauthor
J.P. Calderón,$^{1,2}$
F. Herpich,$^{10}$
E. Telles,$^{11}$ 
K. Saha,$^{12}$
P.A.A. Lopes,$^{3}$
V.H. Lopes-Silva,$^{3}$
T.S. Gon\c{c}alves,$^{3}$
\newauthor
D. Bambrila,$^{3}$
N.M. Cardoso,$^{13}$
M.L. Buzzo,$^{14,15}$
P. Astudillo Sotomayor,$^{16}$
R. Demarco,$^{17}$
N. Leigh,$^{16,18}$
\newauthor
M. Sarzi,$^{19}$
K. Menéndez-Delmestre,$^{3}$
F.R. Faifer,$^{1,2}$
Y. Jim\'enez-Teja,$^{20}$
M. Grossi,$^{3}$
J.A. Hernández-Jiménez,$^{21}$  
\newauthor
A.C. Krabbe,$^{7}$
L.A. Guti\'errez Soto,$^{1}$
D. Brandão,$^{3}$
L. Espinosa,$^{7}$
D. E. Olave-Rojas,$^{22}$
G.B. Oliveira Schwarz,$^{7,23}$
\newauthor
 F. Almeida-Fernandes,$^{7}$
W. Schoenell,$^{24}$
T. Ribeiro,$^{25}$
A. Kanaan,$^{5}$
C. Mendes de Oliveira,$^{7}$\\
\\
$^{1}$Instituto de Astrofísica de La Plata, CONICET-UNLP, Paseo del Bosque s/n, B1900FWA, Argentina\\
$^{2}$Facultad de Ciencias Astronómicas y Geofísicas, Universidad Nacional de La Plata, Paseo del Bosque s/n, B1900FWA, Argentina\\
$^{3}$Valongo Observatory, Federal University of Rio de Janeiro, Ladeira Pedro Antonio 43, Saude Rio de Janeiro, RJ, 20080-090, Brazil\\
$^{4}$Instituto de Fisica, Universidade Federal do Rio de Janeiro, 21941-972, Rio de Janeiro, RJ, Brazil\\
$^{5}$Departamento de Física - CFM - Universidade Federal de Santa Catarina, PO BOx 476, 88040-900, Florianópolis, SC, Brazil\\
$^{6}$Centro Brasileiro de Pesquisas Físicas, Rua Dr. Xavier Sigaud 150, CEP 22290-180, Rio de Janeiro, RJ, Brazil\\
$^{7}$Universidade de São Paulo, IAG, Rua do Mat\~ao 1226, Sao Paulo, SP, Brazil\\
$^{8}$Universidad de Buenos Aires, Facultad de Ciencias Exactas y Naturales y Ciclo Básico Común, Buenos Aires, Argentina\\
$^{9}$CONICET-Universidad de Buenos Aires, Instituto de Astronomía y Física del Espacio (IAFE), Buenos Aires, Argentina\\
$^{10}$Cambridge Survey Astronomical Unit (CASU), Institute of Astronomy, Madingley Road, Cambridge CB3 0HA, UK\\
$^{11}$Observatório Nacional, Rua General José Cristino, 77, São Cristóvão, 20921-400 Rio de Janeiro, RJ, Brazil\\
$^{12}$Inter-University Centre for Astronomy and Astrophysics, Ganeshkhind, Post Bag 4, Pune 411007, India\\
$^{13}$Escola Politécnica, Universidade de São Paulo, Av. Prof. Luciano Gualberto, travessa do politécnico, 380\\
$^{14}$ Centre for Astrophysics and Supercomputing, Swinburne University, John Street, Hawthorn VIC 3122, Australia\\
$^{15}$ARC Centre of Excellence for All Sky Astrophysics in 3 Dimensions (ASTRO 3D), Australia\\
$^{16}$Departamento de Astronomía, Facultad de Ciencias Físicas y Matemáticas, Universidad de Concepción, Concepción, 4030000, Chile\\
$^{17}$Instituto de Astrofísica, Facultad de Ciencias Exactas, Universidad Andr\'es Bello, Sede Concepci\'on, Talcahuano, Chile\\
$^{18}$Department of Astrophysics, American Museum of Natural History, Central Park West and 79th Street, New York, NY 10024, USA\\
$^{19}$Armagh Observatory and Planetarium, College Hill, Armagh BT61 9DG, UK\\
$^{20}$Instituto de Astrof\'isica de Andaluc\'ia--CSIC, Glorieta de la Astronom\'ia s/n, E--18008 Granada, Spain\\
$^{21}$Universidade do Vale do Paraíba, Av. Shishima Hifumi, 2911, São José dos Campos, SP, 12244-000, Brazil\\
$^{22}$Departamento de Tecnologías Industriales, Facultad de Ingeniería, Universidad de Talca, Los Niches km 1, Curicó, Chile\\
$^{23}$Universidade Presbiteriana Mackenzie, R. da Consolação, 930 - Consolação, São Paulo, Brazil\\
$^{24}$GMTO Corporation 465 N. Halstead Street, Suite 250 Pasadena, CA 91107, USA \\
$^{25}$Rubin Observatory Project Office, 950 N. Cherry Ave., Tucson, AZ 85719, USA 
}

\date{Accepted 2024 March 14. Received 2024 March 01; in original form 2023 September 29}
\pubyear{2024}
\begin{document}
\label{firstpage}
\pagerange{\pageref{firstpage}--\pageref{lastpage}}
\maketitle

\begin{abstract}
The Fornax galaxy cluster is the richest nearby (D $\sim$ 20 Mpc) galaxy association in the southern sky. As such, it provides a wealth of oportunities to elucidate on the processes where environment holds a key role in transforming galaxies. Although it has been the focus of many studies, Fornax has never been explored with contiguous homogeneous wide-field imaging in 12 photometric narrow- and broad-bands like those provided by the Southern Photometric Local Universe Survey (S-PLUS). In this paper we present the S-PLUS Fornax Project (S+FP) that aims to comprehensively analyse the galaxy content of the Fornax cluster using S-PLUS. Our data set consists of 106 S-PLUS wide-field frames (FoV$\sim$1.4$\times$1.4 deg$^2$) observed in five SDSS-like {\it ugriz} broad-bands and seven narrow-bands covering specific spectroscopic features like [OII], CaII\,H+K, H$\delta$, G-band, Mg\,b triplet, H$\alpha$, and the CaII triplet. Based on S-PLUS specific automated photometry, aimed at correctly detecting Fornax galaxies and globular clusters in S-PLUS images, our dataset provides the community with catalogues containing homogeneous 12-band photometry for $\sim 3\times10^6$ resolved and unresolved objects within a region extending over $\sim 208$ deg$^2$ ($\sim 5~R_{vir}$ in RA) around Fornax’ central galaxy, NGC\,1399. We further explore the {\sc eagle} and {\sc IllustrisTNG} cosmological simulations to identify 45 Fornax-like clusters and generate mock images on all 12 S-PLUS bands of these structures down to galaxies with  $M_\star \geqslant 10^8~{\rm{M}}_\odot$.  The S+FP dataset we put forward in this first paper of a series will enable a variety of studies some of which are briefly presented. 
\end{abstract}


\section{Introduction}
\label{Intro}
Large astronomical surveys have proven to be natural gateways to new and revolutionary discoveries and invaluable sources of colossal amounts of information. Since the start of the Palomar Observatory Sky Surveys (POSS\,I and POSS\,II) in the 1950's, that imaged the celestial sphere in the B, R and I (POSS\,II) bands with photographic plates, this kind of development keeps on growing spanning all wavelengths from $\gamma$- (e.g.  H.E.S.S., \citealt{HESS}; EGRET, \citealt{Hartman1999EGRET}; GLAST, \citealt{Glast1999}; FERMI-LAT, \citealt{Lott2010Fermi}) and X-rays (e.g. ROSAT, \citealt{ROSAT1981}; X-MM, \citealt{Bleeker1984}; Chandra, \citealp{Hornschemeier2000}) to sub-mm \citep{Campbell1984,Smail1997} and radio regimes \citep{Brown1953,Barnes1997,Waugh2002,Serra2016}. The Sloan Digital Sky Survey (SDSS; \citealt{York2000}), a pioneering photometric and spectroscopic modern survey in the sense that it was performed using CCDs and with a 2.5-m telescope, has driven an essential change in the way to tackle different subjects. Studies previous to SDSS involved the analysis of hundreds or thousands of objects. This paradigm changed, and the study of hundreds of thousands and millions of sources can be addressed. Although the main original goal of SDSS was to build a 3D map of the large-scale structure of the Universe \citep{Bahcall1995}, after 25 years of operation, it is still thrusting the frontiers of both the galactic and extragalactic research fields \citep{almeida2023eighteenth}. 

Wide-field all-sky photometric and spectroscopic surveys are particularly relevant to study the most nearby and rich galaxy clusters which can span a sky projected area of more than 10 deg$^2$ (see, for example, figure\,1 in \citealt{Ferrarese2012} in the case of the Virgo cluster). Thanks to their broad field of view and sky coverage, those surveys allow the exploration of not only the cluster structure up to several times the virial radius (R$_{vir}$), but also their surroundings where the filament structures feeding them reside and where a high rate of galaxy pre-processing is expected \citep{1998Zabludoff,2019Sarron,Kuchner2022}. The paradigmatic cases of this kind of environment are the Virgo cluster ($D=16.5\pm0.1$ Mpc, \citealt{Mei2007})  in the northern sky and the Fornax cluster ($D=20$ Mpc, \citealt{Blakeslee2009}) in the southern hemisphere. 

As a general rule, studies focused on the Fornax cluster have followed those performed on the Virgo cluster in time. The discovery of dwarf galaxies in Fornax by \citet{Hodge1959}, three years after a similar population of objects was reported in Virgo \citep{Reaves1956}, represents a good example of that. The Fornax Cluster Catalogue (FCC, \citealt{Ferguson1989,Ferguson1997yCat}) was compiled from the visual inspection of 26 deep, large-scale plates taken with the du Pont 2.5-m telescope at Las Campanas Observatory four years later than the Virgo Cluster Catalogue (VCC, \citealt{Binggeli1985}), built in a similar manner using plates of the same telescope. Also, the globular cluster systems of the brightest galaxies in Fornax were detected by \citet{Dawe1976} two decades after the identification of a similar system around M\,87 \citep{Baum1955,1956Baum}. 

Modern surveys exploring the Virgo and Fornax clusters are not the exception to that rule. H\,I surveys in Virgo were first carried out in the 1980s \citep{VanGorkom1986,Hoffman1987} while the first systematic exploration of the content of H\,I in the Fornax region was made a decade later \citep{Barnes1997}. X-ray observations of Virgo started in the 1970s \citep{Kellog1971,Kellog1975,Griffiths1974,Forman1979} with the execution of the first dedicated X-ray survey in the 1980s \citep{Forman1985}. In contrast, X-ray observations of Fornax were first reported by \citet{Mason1985}, and the first X-ray survey focused on this cluster was carried out with Chandra in the 2000s \citep{Scharf2005S}. The ACS Virgo Survey \citep{ACS_Virgo2004} started several years earlier than the ACS Fornax Survey \citep{Jordan2007}, a similar situation to that of the Next Generation Virgo \citep{Ferrarese2012} and Next Generation Fornax \citep{Munioz2015} Surveys. 

Being the second nearest rich cluster, Fornax represents a remarkable environment where the processes involved in the formation and evolution of galaxies can be studied in detail. In this respect, Fornax is a dynamically and structurally complex system, with two main components, likely in the process of merging, accompanied by the presence of other substructures, and galaxies and groups being accreted from surrounding filaments \citep{Drinkwater2001,Ordenes2018,Maddox2019,Spavone2022}. Overall, the cluster is dynamically young and still assembling, making it an ideal laboratory to identify and study the effect of different environmental processes on various galaxy types. There is evidence for cold substructures in its main core, which may have resulted from the environmental interactions of galaxies, some of them contributing even to the growth of the central cluster galaxy via mergers \citep{Napolitano2022}. In addition to this, there is currently unambiguous evidence of cluster galaxies being affected by ram-pressure stripping \citep{Zabel2019,Serra2023}, which adds to the observations for pre-processing in one of the main cluster components, Fornax A \citep{Kleiner2021}. The first lists of galaxies in the Fornax region were compiled by \citet{Baker1933} and \cite{Baker1937}. The subsequent comparison of the population of dwarf galaxies in Fornax and Virgo was reported as the first evidence that the evolutionary age of Fornax is greater than that of Virgo \citep{Reaves1964}. Today, about a thousand cluster members are known in Fornax, and many surveys have contributed with comprehensive spectrophotometric data sets from the X-rays to the radio domain \citep{Scharf2005S,Jordan2007,Munioz2015,Iodice2016,Iodice2017,Pota2018,Serra2023}. 

The Southern Photometric Local Universe Survey (S-PLUS) is a 12-band optical survey conducted using the T80-South 0.826-m robotic telescope (T80S) located at Cerro Tololo Interamerican Observatory. The telescope is equipped with a 9.2k $\times$ 9.2k e2v detector with 10-micron pixels, resulting in a field-of-view (FoV) of 2 deg$^2$ with a plate scale of 0.55 arcsec pixel$^{-1}$.  S-PLUS uses the Javalambre 12-band magnitude system \citep{Cenarro2019}, which includes the five $ugriz$ broad-band and seven narrow-band filters centred on prominent stellar spectral features: the Balmer jump/[OII]($\lambda \lambda$3727), CaII H+K, H$\delta$, G-band, Mg b triplet, H$\alpha$, and the CaII triplet. S-PLUS comprises a Main Survey (MS) and four additional sub-surveys with different astrophysical objectives. In particular, the MS aims to cover 9,300 square degrees of the southern sky under photometric conditions and is motivated by extragalactic science  \citep{MendesdeOliveira2019}. 
  
The main goal of this paper is to introduce the S-PLUS Fornax Project (S+FP) aimed at performing a comprehensive study of the Fornax galaxy cluster over 208 square degrees, using homogeneous data taken through the 12 optical bands of S-PLUS. In its first stage, the S+FP consists of 106 S-PLUS contiguous fields around NGC\,1399, all of them observed as part of the S-PLUS MS. This set of fields extends up to $\sim 5$ R$_{vir}$  in right ascention (RA). It is the first time either of the the two nearest rich clusters is analyzed simultaneously in 12 photometric bands and with such a significant sky coverage. 

At the redshift of the Fornax cluster ($z=0.0048$ considering $<V_r>=1442$\,km\,s$^{-1}$ around NGC\,1399; \citealt{Maddox2019}), the astrophysical lines traced by the narrow-band filters of S-PLUS, if present in Fornax members, will fall within the filters, making this data set particularly useful for several studies. Here, we present the initial galaxy sample to be analyzed in the framework of the S+FP and describe some preliminary results of different ongoing studies in several areas such as automatic photometry, identification of new cluster members, H$\alpha$ emitters and peculiar objects like blue elliptical galaxies, spectral energy distribution (SED) fitting analysis, morphological and stellar population characterization using Machine and Deep Learning techniques, detection of globular clusters and comparison of the observational results with the properties of Fornax-like clusters identified in numerical simulations. 

The paper is organized as follows. In Section\,\ref{sec:data} we present a compilation of Fornax galaxies reported in the literature, the S+FP data and the initial galaxy sample to be analyzed. In Section\,\ref{sec:results}, we present preliminary results on the different topics covered by the S+FP, and in Section\,\ref{sec:conclusions}, we give our concluding remarks. AB magnitudes are used throughout this paper. In addition, we will assume a distance modulus of $(m-M)=31.51$ mag for Fornax \citep{Blakeslee2009} and, at the corresponding distance, 1 arcsec subtends 0.1 kpc. 

\section{Data}
\label{sec:data}
\subsection{Fornax Literature Galaxy Sample}
\label{subsec:literature}
The galaxy content of the Fornax cluster has been extensively studied since the seminal papers by \citet{Hodge1959,Hodge1960}, which provided the initial compilation and analysis of dwarf galaxies. Our first step in the framework of the S+FP was to set a reference sample of Fornax galaxies reported in the literature as spectroscopically confirmed members or likely members according to morphological criteria (e.g. objects displaying dwarf-like shapes or low surface brightnesses in photographic plates or CCD images; \citealt{Ferguson1989,Munioz2015,Venhola2017}). To achieve that goal,  we conducted an extensive search of NASA's Astrophysics Data System (ADS) for references related to galaxy catalogues in the Fornax cluster region. From that search, we ended up with a list of 1,057 objects reported in 21 catalogues of galaxies  \citep{Hodge1965,Davies1988,Ferguson1989,Irwin1990,Schroeder1996,Graham1998,Hilker1999b,Morshidi1999,Drinkwater2001,Karick2003,Bergond2007,Mieske2007,Jordan2007,Munioz2015,Kourkchi2017,Venhola2017,Eigenthaler2018,Ordenes2018,Venhola2018,Raj2019,Maddox2019}. 
For all those galaxies, we checked the existence of radial velocities in the NASA/IPAC Extragalactic Database (NED), and we visually inspected their morphologies using the Legacy Surveys Sky Viewer\footnote{\url{https://www.legacysurvey.org/viewer}}. That helped us to remove background galaxies from our list. The final compilation of spectroscopically confirmed or likely galaxy members of the Fornax cluster includes 1,005 objects of which only 233 (23\%) have reported radial velocities. We will refer to this sample of galaxies as the Fornax Literature Sample (FLS).

\subsection{Fornax S-PLUS images}
\label{subsec:images}
\begin{figure}
\centering
\includegraphics[width=1.0\columnwidth]{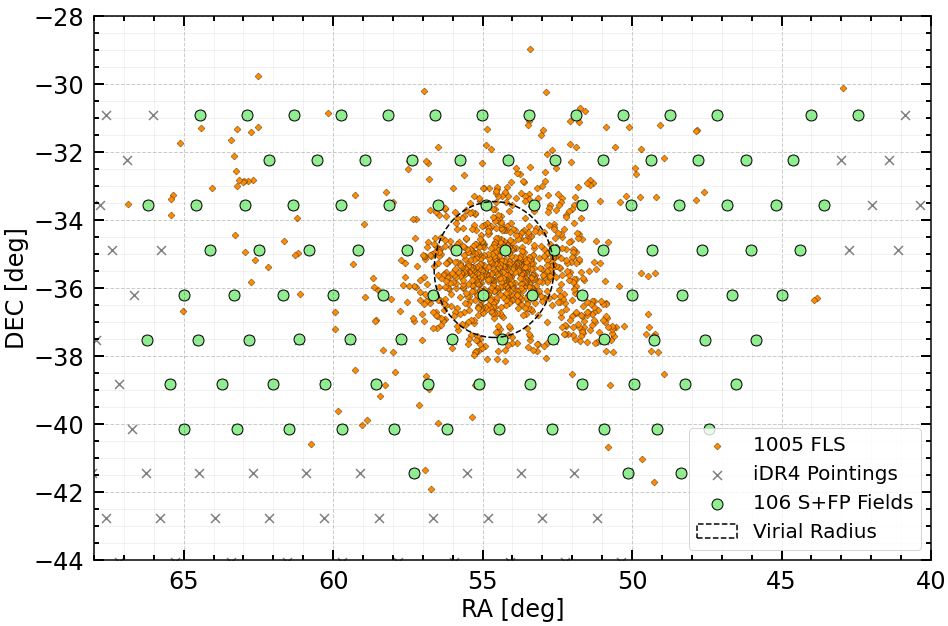}
 \caption{Spatial distribution of the 106 S+FP fields and the Fornax Literature Sample (FLS).}
 \label{pointings}
\end{figure}

In the context of the S+FP, we are analyzing 106 S-PLUS fields observed as part of the publicly available S-PLUS Data Release 3 (DR3; 43 fields) and the S-PLUS internal DR4 (iDR4; 63 fields). They cover a sky region of $\sim$ 23 $\times$ 11 deg$^2$, including 1,000 galaxies from the FLS and extending to the outskirts of the Fornax group up to approximately 5 R$_{vir}$ (see Figure\,\ref{pointings}). Each field was observed in the 12 photometric bands of S-PLUS.  
The observations were done during dark-grey time and follow an east-west dithering pattern to mitigate the effects of bad columns in the detector. In that way, it is possible to use objects located through the entire CCD area without a big impact caused by instrumental defects. For more details on the telescope and the instrumentation, we refer the reader to \citet{MendesdeOliveira2019}. 

\begin{figure*}
\centering
\includegraphics[width=0.934\columnwidth]{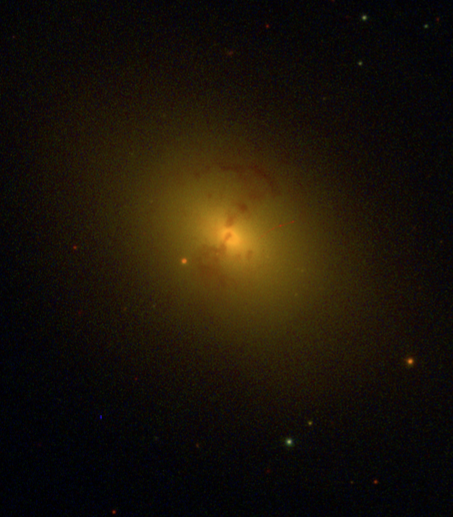}
\includegraphics[width=1.066\columnwidth]{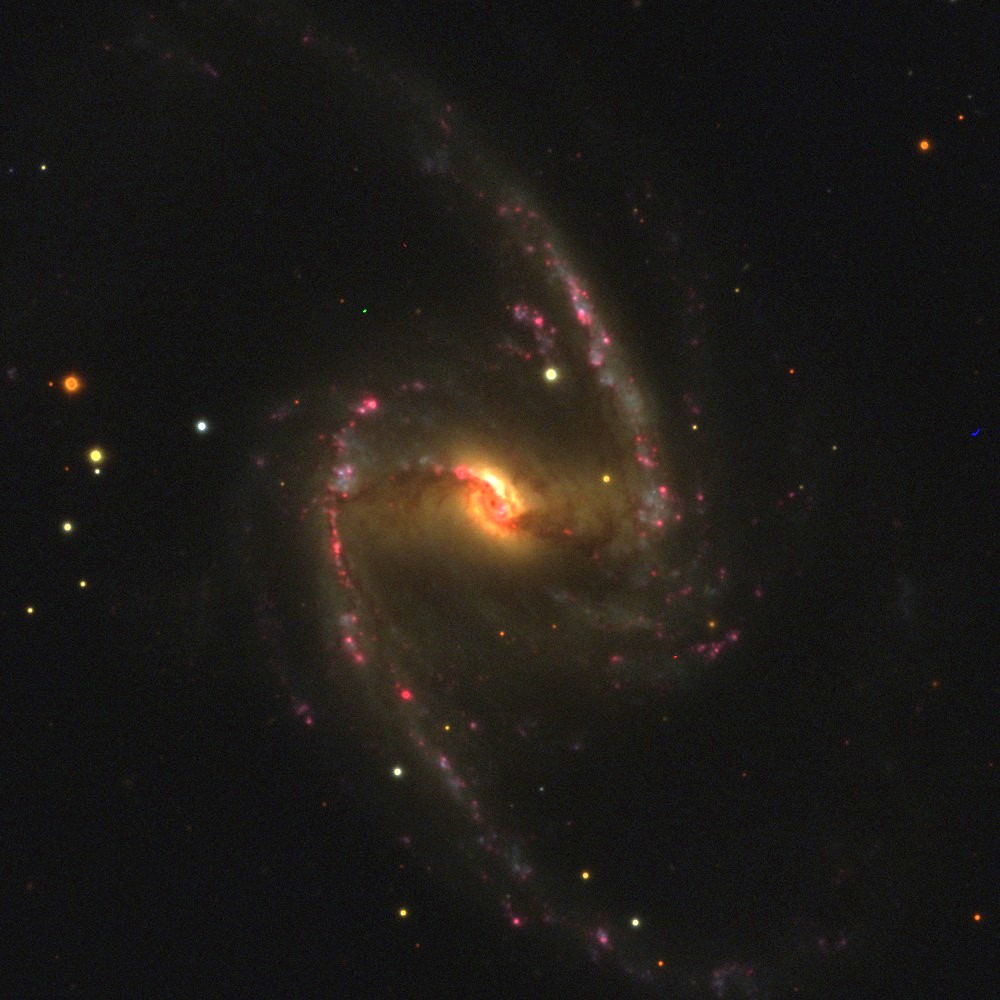}
 \caption{RGB S-PLUS composites of NGC\,1316 {\it (left)} and NGC\,1365 {\it (right)}, two of the brightest galaxies of the Fornax cluster, built following \citet{Lupton2004}. The colour code of these composites (R=$J0660$, G=$g$ and B=$u$+$J0378$) is chosen to highlight H$\alpha$+[NII] regions in red.}
 \label{lupton}
\end{figure*}

In order to show the potential of combining S-PLUS broad- and narrow-band images, Figure\,\ref{lupton} shows RGB S-PLUS composites of two of the brightest galaxies of the Fornax cluster (NGC\,1316 and NGC\,1365) built following \citet{Lupton2004}. The colour code (R=$J0660$, G=$g$ and B=$u$+$J0378$) is chosen to highlight H$\alpha$+[NII] regions in red, as it can be seen in the case of NGC\,1365. It is also remarkable how the inner structure of NGC\,1316 is recovered by the S-PLUS filters, specially taking into account the resolution and depth of the survey. 

\subsection{Fornax S-PLUS catalogues}
\label{subsec:catalogs}

\begin{figure*}
 \includegraphics[width=2.0\columnwidth]{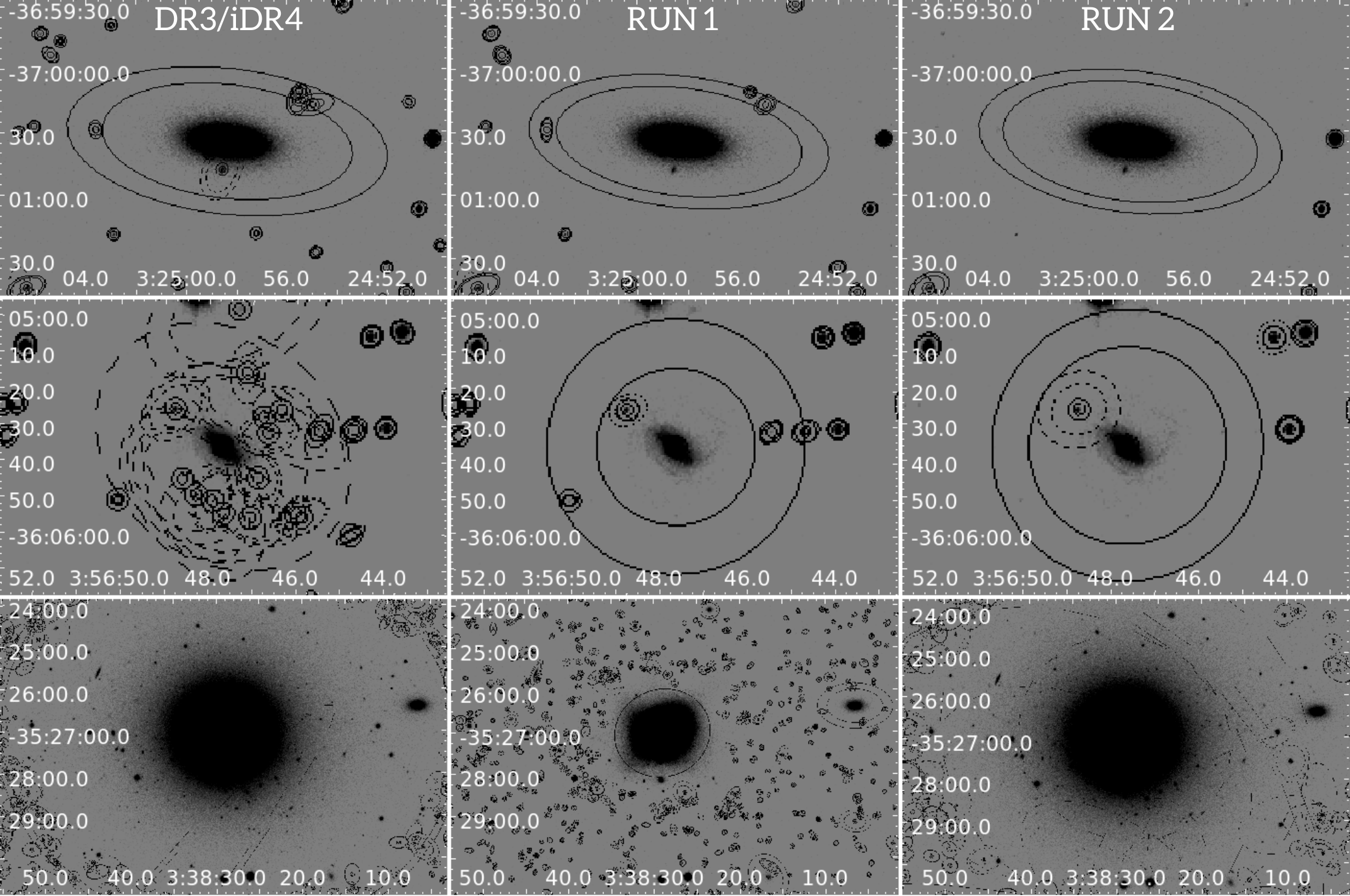}
 \caption{{\it Left:} Aperture images obtained using \texttt{SExtractor} input parameters for S-PLUS DR3 and iDR4. {\it Center:} Aperture images obtained using RUN\,1 \texttt{SExtractor} input parameters. {\it Right:} Aperture images obtained from RUN\,2 \texttt{SExtractor} input parameters. All panels in the same line show the same object with the same spatial and brightness scale.}
 \label{comparacion}
\end{figure*}

Besides the 12-band images for each observed field, S-PLUS delivers catalogues with homogeneous 12-band photometry for all the objects detected in the fields. As a starting point, in the context of the S+FP, we worked with the DR3 and iDR4 S-PLUS catalogues. Those catalogues are obtained using \texttt{\texttt{SExtractor}}\footnote{\url{https://www.astromatic.net/software/SExtractor}} \citep{Bertin1996} with specific input parameters common to all S-PLUS releases and chosen to provide valuable data to both galactic and extragalactic topics. In other words, these parameters are not optimized for detecting any specific type of astronomical object. After making a cross-match between those catalogues and the FLS, and after inspecting the \texttt{SExtractor} aperture images obtained by S-PLUS for each field, we noticed that several Fornax galaxies were missed or excessively deblended in the S-PLUS DR3 and iDR4 catalogues. In the left panels of Figure\,\ref{comparacion}, we show several examples of that situation.

\begin{figure*}
        \centering
        \includegraphics[width=1.0\textwidth]{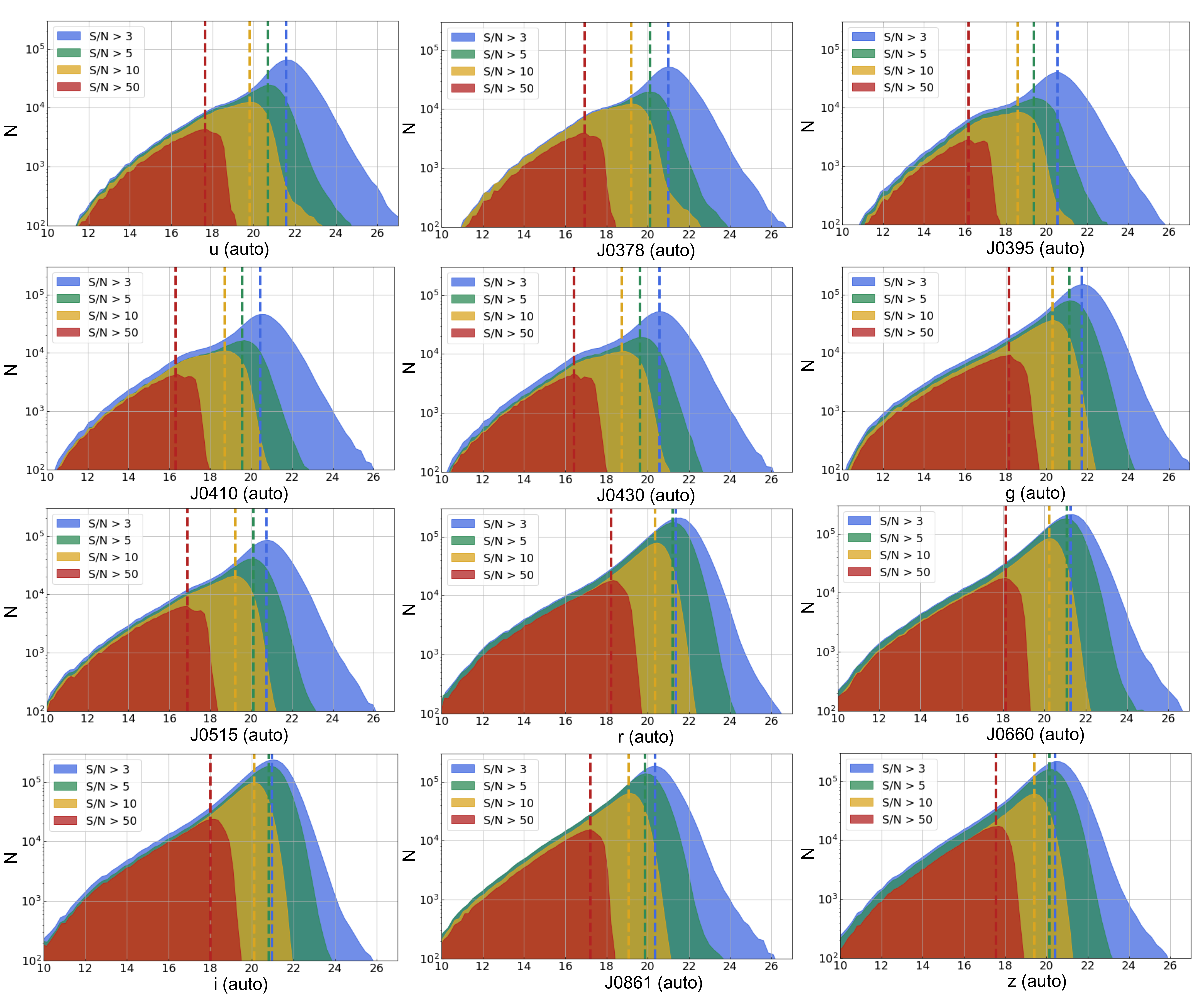}
\caption{Photometric depth of RUN\,1 in the 12 bands of S-PLUS considering four signal-to-noise (S/N) thresholds (S/N > 3, blue; S/N > 5, green; S/N > 10, yellow; S/N > 50, red) for each filter. In all cases, we consider S/N=5000 as a maximum acceptable S/N which corresponds to the S/N displayed by NGC\,1399 in the $i$-band. The characteristic photometric depth at each S/N threshold and for each filter is given by the peak of the magnitude distributions (dashed coloured lines).}
\label{completitud}
\end{figure*}

To automatically measure confidently as many galaxies from the FLS as possible, we run \texttt{SExtractor} with two different sets of parameters optimized to detect faint objects near bright galaxies (hereafter, RUN\,1) and bright and large galaxies with apertures that correctly estimate their sizes (hereafter, RUN\,2). The central and right panels of Figure\,\ref{comparacion} show the results obtained from RUN\,1 and RUN\,2 compared to DR3 and iDR4 runs. In Table\,\ref{tab:Parameters}, we compare the input parameters of \texttt{SExtractor} used to obtain the photometry of DR3 and iDR4, and those considered in RUN\,1 and RUN\,2. The RUN\,1 catalogue includes 2,900,926 objects, while that obtained using the RUN\,2 parameters contains 1,390,237 sources. Following the analysis performed by \citet{Almeida-Fernandes2022} for the DR2 of S-PLUS, in Figure\,\ref{completitud} we show the photometric depth in the 12 bands of S-PLUS in the case of RUN\,1, which contains the faintest and most compact objects of both samples. For details about the implementation and performance of those runs, we refer the reader to Haack et al. (submitted). 

\begin{table}
\centering
\caption{Comparison of the input parameters of \texttt{SExtractor} to obtain the catalogues of DR3 and iDR4 \textit{(left)}, RUN\,1 \textit{(centre)} and RUN\,2 \textit{(right)}}
\label{tab:Parameters}
\begin{tabular}{|c|c|c|c|}
\hline
Parameters & DR3/iDR4 & RUN\,1 & RUN\,2 \\
\hline
 DETECT\_MINAREA & 4 & 5 & 10 \\ 
 DETECT\_THRESH & 1.1 & 1.5 & 1.5 \\  
 ANALYSIS\_THRESH & 3. & 1.5 & 1.5 \\
 FILTER & Y & Y & Y \\ 
 FILTER\_NAME & tophat\_3\_3x3 & default & gauss\_5\_9x9 \\  
 DEBLEND\_NTHRESH & 64 & 32 & 64 \\
 DEBLEND\_MINCONT & 0.0002 & 0.005 & 0.001 \\ 
 BACK\_SIZE & 256 & 64 & 524 \\  
 BACK\_FILTERSIZE & 7 & 3 & 3 \\
 BACKPHOTO\_TYPE & LOCAL & LOCAL & GLOBAL \\
 BACKPHOTO\_THICK & 48 & - & - \\
 BACK\_TYPE & - & - & MANUAL \\
 BACK\_VALUE & - & - & 0.0 \\
 \hline
\end{tabular}
\end{table}

As the initial step in creating a master catalogue of galaxies with homogeneous S-PLUS photometry, we refined the catalogues obtained from RUN\,1 and RUN\,2. We applied the following criteria:

\begin{itemize}
    \item CLASS\_STAR\_{\it gri} < 0.35,
    \item {\it gri}\_AUTO < 22 mag.
\end{itemize}

\noindent The first conditions select objects with simultaneous values of the parameter CLASS\_STAR in the {\it g}, {\it r}, and {\it i} bands in agreement with being extended sources.  The second criterion avoids objects with magnitude values of 99 in those three broad bands. In that way, we obtained two restricted catalogues from RUN\,1 and RUN\,2 containing 382,233 (RUN\,1$_{res}$=13\% RUN\,1) and 380,445 (RUN\,2$_{res}=27$\% RUN\,2) extended sources, respectively. As a second step, we merged the RUN\,1$_{res}$ and RUN\,2$_{res}$ catalogues, removing duplications. In the case of an object was detected both by RUN\,1 and RUN\,2, we keep the measurement from RUN\,2 as it better characterizes the sizes (and, as a consequence, the total brightnesses) of extended objects. In that way, we got a final catalogue (RUN\,1+RUN\,2)$_{res}$ containing 458,932 extended sources that will be the base of all the subsequent analyses. To understand this final number of objects, it is worth noticing that not all objects included in RUN\,2$_{res}$ were properly detected by RUN\,1 and vice versa.

\begin{figure}
    \centering
    \includegraphics[width=0.48\textwidth]{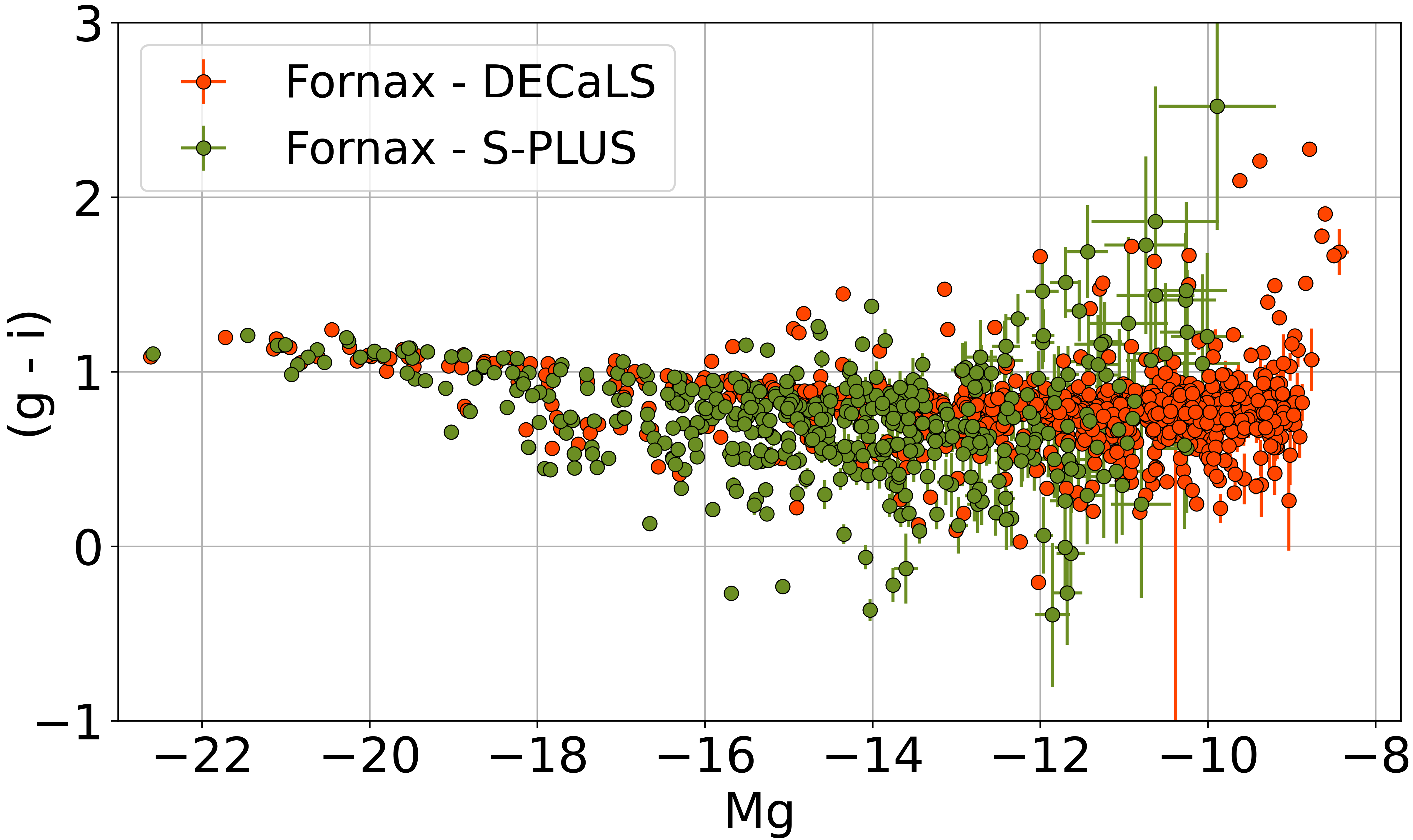}
    \caption{($g - i$) vs. Mg diagram of the S+FP initial galaxy sample (IGS) using the S+FP broad band photometry (green dots). For comparison, we also show the colour and magnitude of the same galaxies but with the photometry provided by the DECam Legacy Surveys (DECaLS) DR10 (orange dots). We can see that the broad-band photometry obtained from S-PLUS is consistent with the deeper photometry of DECaLS, although it has a larger colour dispersion due to larger photometric errors, as expected.}
    \label{CMR_IGS}
\end{figure}

From the cross-match of the FLS with the (RUN\,1+RUN\,2)$_{res}$ catalogue, considering 10 arcsec in the error of the match to avoid losing the most extended galaxies, we got a final list of 443 Fornax literature galaxies with confident S-PLUS photometry. Hereafter, we will refer to that subset of galaxies as our Initial Galaxy Sample (IGS). In Figure\,\ref{CMR_IGS}, we show the location of the S+FP IGS in the ($g - i$) vs M$_g$ diagram. For comparison, we also plot the location of the FLS cross-matched with the DECam Legacy Survey (DECaLS; \citealp{Dey2019}) DR10 catalogues, which contain $griz$ broad-band photometry. Our photometry nicely recovers the colour-magnitude relation of the Fornax cluster and, although displaying a larger colour scatter due to larger photometric errors, in a consistent manner to that obtained from DECaLS photometry. It should be noticed that DECaLS magnitudes are obtained from images taken with the 4-m Blanco telescope and, consequently, are deeper than those obtained by S-PLUS. 
Therefore, this plot affirms the reliability of our data for conducting subsequent analyses. It is also noteworthy that at the faint end (M$_g$ > -11 mag) of the CMR, the IGS encompasses previously reported low surface brightness (LSB) galaxies, which may not have been anticipated given the characteristics of S-PLUS (see Section\,\ref{LSBs}).

\begin{figure*}
    \centering    
    \includegraphics[width=1.0\textwidth]{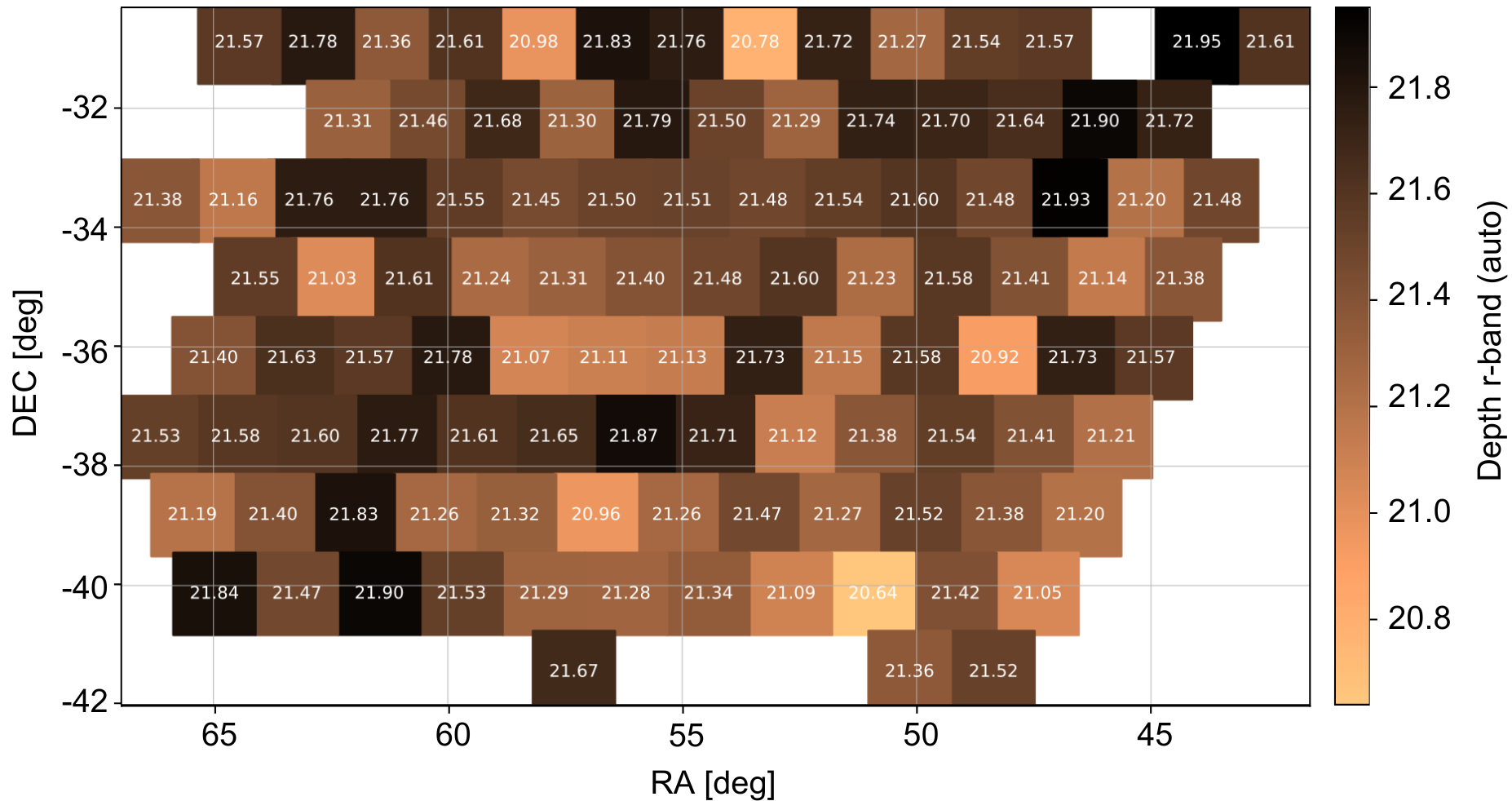}
 \caption{Photometric depth of the 106 S+FP fields in the $r$-band. The values shown in this plot correspond to the AUTO $r$-band magnitude of the faintest object detected in each field by RUN\,1 (S/N > 3), which allows to detect the faintest and most compact objects of the whole S+FP photometric sample.}
\label{Depth}
\end{figure*}

Figure\,\ref{Depth} illustrates the $r$-band photometric depth reached in the 106 S+FP fields. The values in the plot correspond to the AUTO $r$-band magnitude of the faintest object detected using the parameters of RUN\,1 in each field. It is important to note that RUN\,1 is designed to detect the faintest and most compact objects among the two \texttt{SExtractor} runs. Table \ref{tab:statistical_information} shows a summary of the statistical information obtained from the depth map. It can be seen that, although the depth variation between the deepest and the shallowest field is $\sim$ 1.3 mag, the standard deviation is low and similar to an acceptable photometric error ($\sim$ 0.3 mag). The large and asymmetric fluctuation range can be explained by the fact that the images corresponding to the 106 S+FP fields were taken along 22 nights, with different observing conditions of air mass, sky background and transparency.

\begin{table}
\centering
\caption{Statistical information of the photometric depth of the 106 S+FP fields (see Figure\,\ref{Depth}).}
\label{tab:statistical_information}
\begin{tabular}{|c|c|}
\hline
Statistical parameter & Values \\
 & [mag]\\
\hline
 Minimum & 20.64 \\ 
 Maximum & 21.95 \\
 Mean & 21.46 \\ 
 Median & 21.50 \\  
 Standard Deviation & 0.26 \\
 Q$_{10\%}$ & 21.12 \\
 Q$_{1}$ & 21.29 \\
 Q$_{3}$ & 21.63 \\
 Q$_{90\%}$ & 21.78 \\
 \hline
\end{tabular}
\end{table}
\section{Preliminary Results}
\label{sec:results} 
\subsection{Spectroscopic membership and cluster properties}
\label{spec_memb}

We considered Fornax galaxies from the spectroscopic census performed by \citet{Maddox2019} to select members and also estimate cluster properties, such as R$_{vir}$, velocity dispersion ($\sigma_{\text{cl}}$)  and virial mass (M$_{vir}$). We applied the shifting gapper technique \citep{fad96, lop09} to all galaxies with available redshifts to distinguish between Fornax members and interlopers.  From the 236 galaxies identified in the study by \citet{Maddox2019}, we classified 232 as members, with 185 of them falling within the R$_{vir}$. Initially, we estimated $\sigma_{\text{cl}}$ using all member galaxies. Subsequently, we calculated an estimate of M$_{vir}$ using equation 1 from \citet{fer20}, while also applying the corrections recommended by these authors. Further details on these estimates can be found in \citet{lop09,lop14,lop18} and \citet{fer20}. Our results for the Fornax cluster are as follows: $\sigma_{\text{cl}} = 346.4$ km/s, M$_{vir}= 0.49 \times 10^{14}$ M$_{\odot}$, and R$_{vir}=0.74$ Mpc. These values are in good agreement with those reported by \citet{Maddox2019}.

\subsection{Clustering algorithm for membership classification}
\label{sec:camc}

At the Fornax distance, it is not possible to obtain accurate photometric redshifts for Fornax members using S-PLUS as typical photometric redshift errors are 2-3 times greater than the Fornax mean redshift ($\sigma_z=0.02-0.03$, $\langle z\rangle_{\rm Fornax} \sim 0.005$; \citealt{Lima2022}). In that context, and given the large number of extended sources detected and measured by SExctractor over the 106 S+FP fields, it is interesting to test clustering algorithms to find new Fornax members based only on photometric parameters and relative positions within the cluster. Our approach combines Principal Component Analysis (PCA) and a Random Forest classifier (RF). With the former, we characterise the sample of extended sources using the available photometry in the 12 bands, the structural parameters computed by \texttt{SExtractor}, and each source's sky position (RA, DEC). At the same time, the latter is used to classify data and identify previously uncatalogued galaxies within the Fornax cluster, capitalizing on the insights gleaned from PCA's dimensionality reduction and feature extraction.

As it was mentioned in Section\,\ref{subsec:catalogs}, we acquired confident calibrated photometry and structural parameters for $\sim460,000$ extended sources detected in the 106 S+FP fields using \texttt{SExtractor}, all included in the (RUN\,1+RUN\,2)$_{res}$ catalogue. Among them, we found 443 galaxies from the FLS that are refered as IGS  (see Section\,\ref{subsec:catalogs}). 
Of these galaxies, 143 have been spectroscopically confirmed as Fornax members, falling within the velocity range of 600 - 3,000 km\,s$\rm ^{-1}$ \citep{Maddox2019}. This subset of galaxies will be referred to as our Fornax spectroscopic sample (FSS). From a compilation of radial velocities in the covered area\footnote{\url{https://github.com/ErikVini/SpecZCompilation}}, we also set up a sample of 2,986 confirmed background galaxies. We will refer to this sub-sample as our background spectroscopic sample (BSS). We extracted the FSS and BSS from the (RUN\,1+RUN\,2)$_{res}$ catalogue, obtaining a restricted sample (RS) of 59,716 objects without radial velocities.

The method achieves high precision and accuracy, both reaching 0.99, when utilizing the combined spectroscopic samples, FSS + BSS, as the training dataset. To deal with the imbalance between FSS and BSS, we used the Synthetic Minority Oversampling Technique (SMOTE; \citealp{Chawla2011}). It operates by randomly selecting a data point from the minority class and determining the k-th nearest neighbours for that point. Synthetic points are generated and inserted between the selected point and its neighbours. 

\begin{figure}
    \centering
    \includegraphics[width=1.0\columnwidth]{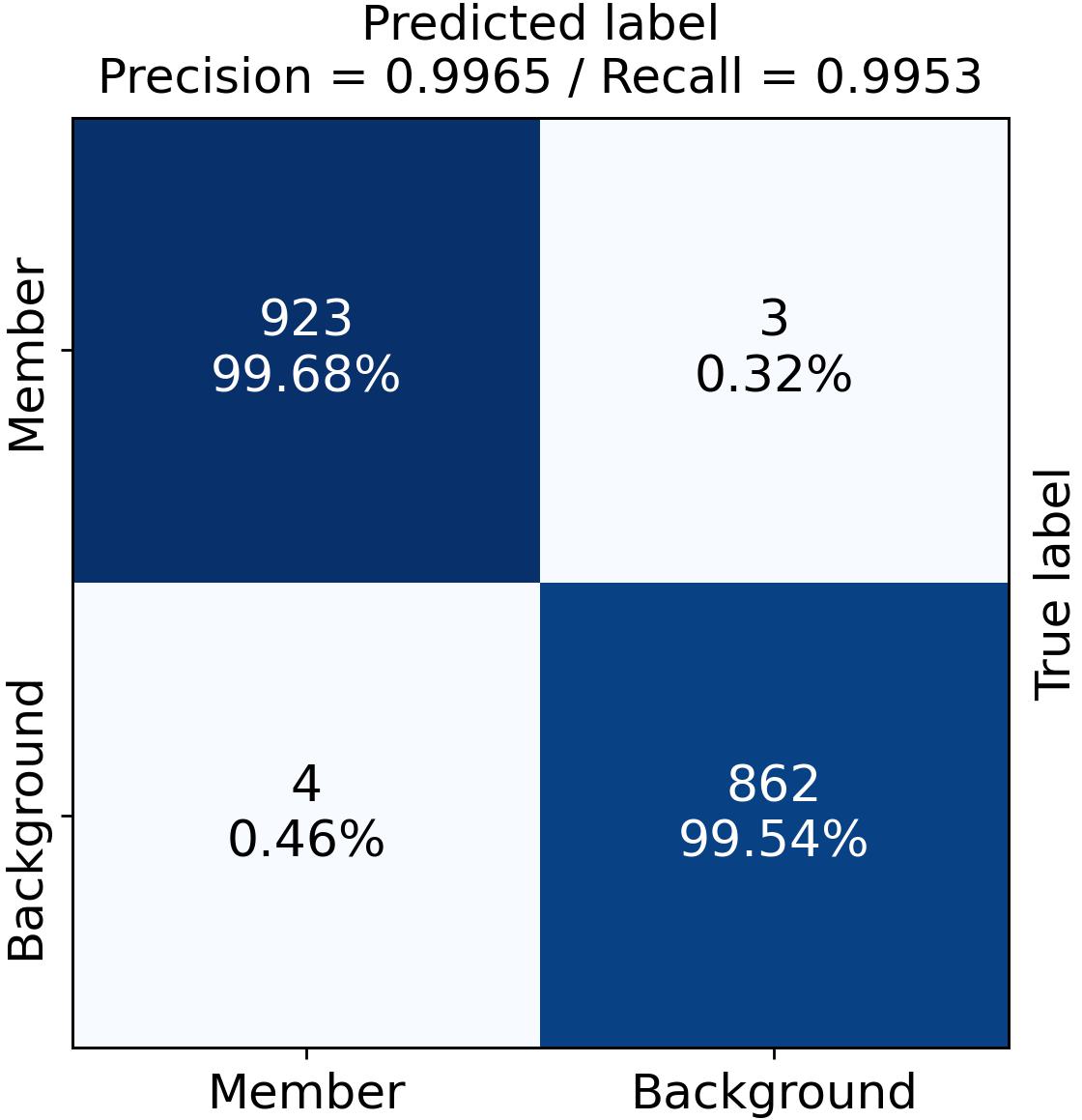}
 \caption{Confusion matrix resulting from applying the Random Forest algorithm on the Fornax and background spectroscopic samples of Section\,\ref{sec:camc}. }\label{Confusion_matrix}
\end{figure}

We can see the confusion matrix in Figure\,\ref{Confusion_matrix}, which demonstrates that the classifier effectively distinguishes cluster members from background sources with a high level of confidence. The result of our test for clustering membership is that 6,470 objects of the RS have a probability (p) larger than 0.5 of being a Fornax cluster member. 

\begin{figure}
    \centering
    \includegraphics[width=0.49\columnwidth]{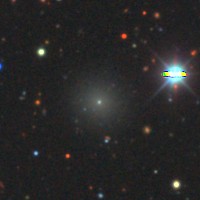}
    \includegraphics[width=0.49\columnwidth]{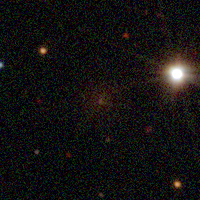}\\
    \includegraphics[width=0.49\columnwidth]{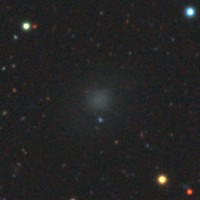}
    \includegraphics[width=0.49\columnwidth]{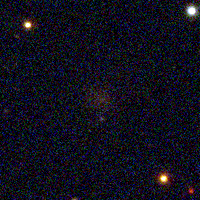}
    \caption{DECaLS {\it(left)} and S-PLUS {\it(right)} RGB images of two examples of FS89 galaxies with membership status 1 and identified by our classifier as objects with p $\leq$ 0.5 of being Fornax cluster members. Although LSB galaxies are detected in the S-PLUS images (see Section\,\ref{LSBs}), our identification method is not capable of selecting them as Fornax members due to their low S/N.}
    \label{prob_less_0.5}
\end{figure}

\begin{figure}
    \centering   
    \includegraphics[width=1.0\columnwidth]{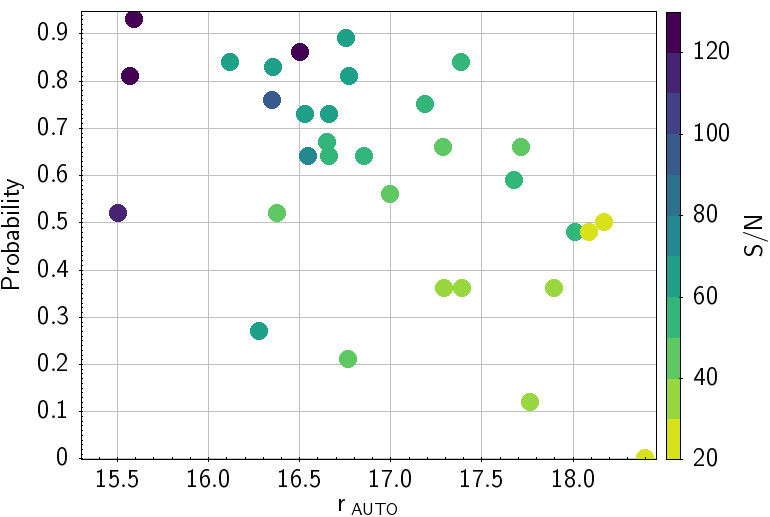}
    \caption{Probability of being a Fornax cluster member versus $r$-band AUTO magnitudes, colour coded according to the S/N in the $r$-band, for galaxies with Fornax membership status 1 (confident member) and 2 (likely member) assigned by FS89 in morphological basis. For this specific set of galaxies (32 objects), there seems to be a correlation between the probability of being a Fornax cluster member assigned by our method, and the brightness and S/N displayed by the objects in the S-PLUS images.}
    \label{S/N_FS90}
\end{figure}

To validate our results, we performed a cross-match considering a coordinate offset of 10 arcsec, between the RS and the Fornax catalogue obtained by \citet[][hereafter, F89]{Ferguson1989}. Each galaxy included in the F89 catalogue is assigned a membership status based on morphological criteria: status 1 galaxies are considered as confident members, status 2 galaxies are likely members and status 3 galaxies are possible (unlikely) members. In this cross-match, we identified 32 F89 galaxies within the RS. Among them, 25 have a membership status 1 and seven were assigned a membership status 2. If we take a look at the probability values assigned by our method to the 32 F89 galaxies in the RS, 18 of the 25 galaxies with status 1 (72\% of the status 1 sample) and four of the seven galaxies with status 2 (57\% of the status 2 sample) have p > 0.5. If we consider the complete set of status 1+2 galaxies in the RS (32 objects), we found that 22 galaxies (69\% of the status 1+2 sample) have p > 0.5. The remaining 10 F89 galaxies (31\% of the status 1+2 sample), which display p $\leq$ 0.5, are LSB galaxies (see Figure\,\ref{prob_less_0.5} for some examples). That might indicate that our procedure for identifying Fornax members would be confident for galaxies displaying a brightness above a certain threshold. This is confirmed by Figure\,\ref{S/N_FS90} that shows what seems to be a correlation between the probability of being a Fornax cluster member assigned by our method, and the $r$-band AUTO magnitude and the S/N displayed by the object in the $r$-band image. Additional analysis of these results and an extensive explanation of the procedure will be presented in Calderón et al. (in prep.).

\subsection{Luminosity Function}

We have fitted a single Schechter function to the magnitude distribution of the IGS (see Section\,\ref{subsec:catalogs}) using the maximum likelihood estimator proposed by \cite{Sandage_Tamman_1979} (see also \citealt{Willmer1997}), with errors estimated by a bootstrap methodology.
We do not observe any significant dip at magnitudes fainter than the characteristic magnitude $M^*$, as observed in some studies (e.g., \citealt{Loveday2012} for the field luminosity function). Additionally, the magnitude interval of the sample analyzed here is not very large (see below). These two considerations justify using a single Schechter function to model the cluster luminosity distribution.

\begin{figure}
  \begin{center}
 \includegraphics[width=1.0\columnwidth,angle=0]{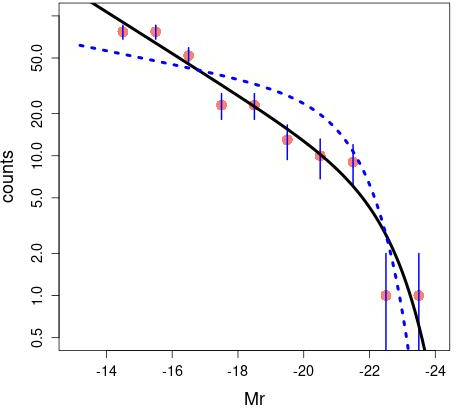}
\caption{S-PLUS Fornax luminosity function (in black, arbitrary units) for galaxies brighter than $M_{r_{petro}} = -14$ mag. The dots with error bars are counted in bins with poissonian errors. The blue dashed line is the luminosity function (arbitrary units) derived for the GAMA survey by \citet{Kelvin2014}.}
\label{fig-lumfun16}
  \end{center}
  \end{figure}

For this exercise, we considered galaxies brighter than $r_{petro} = 17.5$ in the IGS since that sample  appears to be complete up to this magnitude limit. Assuming a distance modulus of 31.51 mag for the Fornax cluster \citep{Maddox2019}, this magnitude limit corresponds to $M_{r_{petro}} \simeq -14$ mag. Such an IGS sub-sample comprises 286 galaxies, out of which 210 have radial velocities consistent with those of cluster members. The best-fit parameters and their uncertainties (corresponding to a 95\% confidence interval) are $M_{r_{petro}}^\star = -22.8_{-1.0}^{+1.1}$ and $\alpha = -1.37_{-0.07}^{+0.08}$ (the reduced $\chi^2$ is 1.85). This result is shown in Figure\,\ref{fig-lumfun16}, where we plotted our unbinned estimate of the cluster luminosity function together with the binned magnitude distribution of our sample.

These results can be compared with other estimates of the luminosity function for the Fornax cluster and the overall galaxy population at low redshifts. One of the earliest estimates of the Fornax luminosity function was obtained by \citet{Ferguson_Sandage_1988}.  
Transforming their photographic $B_T$ magnitudes to the $r$ band with \citet{Fukugita1995} conversion relations for elliptical galaxies, we get  $M_{r^\prime}^\star \simeq -20.8_{-3.1}^{+2.0}$ and $\alpha = -1.32_{-0.08}^{+0.09}$ with confidence limits corresponding to 70\% (top) and 99\% (bottom) probabilities. Within the errors, these results are consistent with ours. More recently, \citet{Kelvin2014} provided results for the field luminosity function using data from the GAMA survey, reporting $M_r^\star =21.7 \pm 0.1$ and  $\alpha =  -1.12 \pm 0.03$. Compared with these results, our S-PLUS Fornax luminosity function has a brighter $M^\star$ and a steeper slope (see Figure\,\ref{fig-lumfun16}).

\subsection{Emission Line Galaxies}
\label{subsec:elgs}
Emission lines are spectral features essential to identifying different physical processes such as star formation (with a timescale of $\sim$10 Myrs), through H$\alpha$, [OIII] or [OII] \citep[e.g.][]{Kennicutt2012}, or the existence of an active galactic nucleus, using the BPT diagram \citep[e.g.][]{BPT1981} that relates [OIII]/H$\beta$ to [NII]/H$\alpha$, among others. In the specific context of a galaxy cluster, we aim at finding galaxies with these lines, namely Emission Line Galaxies (ELGs). When we combine their presence with other physical properties such as morphological type, stellar mass, and their positions within the cluster, along with HI data, we can gain valuable insights into the cluster's evolutionary history.

At the distance of the Fornax cluster, the [OII] and H$\alpha$+[NII] emission lines fall within the S-PLUS $J0378$ and $J0660$ narrow-band filters, respectively. In a preliminary analysis, the presence of these lines can be identified through an excess in the colours $(u-J0378)$ and $(r-J0660)$, respectively. However, for an in-depth study of these sources, it is essential to correct them for the contribution of the galaxy continuum. For this, we apply the Three Filter Method \citep[3FM, e.g.][]{Pascual2007}, where the primary assumption is that the two broad-band neighbouring filters of a given narrow-band can trace the continuum of a source within the narrow filter. In the case of [OII] emitters, we use information from the $u$, $J0378$ and $g$ bands, while for H$\alpha$+[NII] emitters, we need data from the $r$, $J0660$ and $i$ bands.  

To identify ELGs and characterize their spatial emission distribution, we directly applied the 3FM to the S-PLUS images. For this purpose, we developed a Python code called Pixel-to-pixel Emission Line Estimate (P.E.L.E; Lopes et al., in prep.). P.E.L.E. creates emission line maps by applying the 3FM to each pixel within a set of images. When RA and DEC coordinates are given as input, along with the image size in pixels, the program automatically carries out the following steps:

\begin{itemize}
    \item downloads the three necessary band images from the S-PLUS database;
    \item calibrates each image using the zero-points derived by \cite{Almeida-Fernandes2022};
    \item masks the stars based on an automatic cross-match to GAIA\,DR3 \citep{GaiaCollab2016, GaiaCollab2022};
    \item generates a data cube including a masked star version of each flux-calibrated image;
    \item creates a background-subtracted data cube considering a 
    2D background model for each filter, where the background level in each mesh is the sigma-clipped median with a 3$\sigma$ cut;
    \item removes high spatial-frequency noise by applying Butterworth spatial filtering \citep{Menezes2014} to the data cube;
    \item applies the 3FM equations presented in \cite{VilellaRojo2015} to each unmasked pixel in the data cube, resulting in an emission line flux map.
\end{itemize}

In this subsection, we present preliminary results of this approach to create H$\alpha$+[NII] emission line maps for Fornax members. We intend to deepen our analysis of these maps and explore [OII] emission in a forthcoming paper devoted to this topic (Lopes et al., in prep.).

\begin{figure*}
\begin{center}
\includegraphics[width=1.0\textwidth]{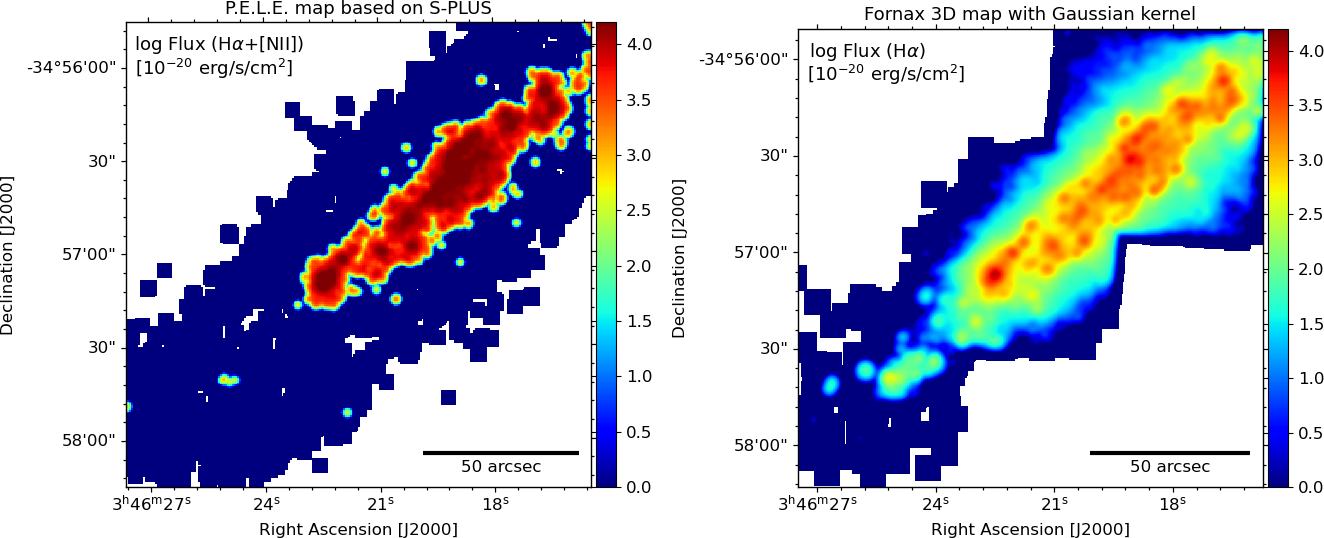}
\caption{Comparison between the H$\alpha$+[NII] map created by P.E.L.E. (left) and the H$\alpha$ map obtained by the Fornax\,3D project (right) for FCC\,312. The panel on the right results from the convolution of the H$\alpha$ map presented in \citet{Iodice2019b} with a Gaussian kernel. The colour scale represents the logarithmic fluxes in units of 10$^{-20}$ erg s$^{-1}$ cm$^{2}$. }
\label{fig.halpha-map-comparison}
\end{center}
\end{figure*}

To validate our approach, we compare our H$\alpha$+[NII] maps to H$\alpha$ maps obtained by integral field spectrograph imaging in the context of the Fornax\,3D project \citep[F3D, ][] {Sarzi2018}. As our technique has a lower resolution than F3D, we apply a Gaussian kernel to the original maps of F3D presented in \cite{Iodice2019b} to match the resolutions. Figure\,\ref{fig.halpha-map-comparison} shows an example of the H$\alpha$+[NII] map derived by P.E.L.E and the F3D H$\alpha$ map for the galaxy FCC\,312. This comparison demonstrates that P.E.L.E, applied to S-PLUS images, is able to recover H$\alpha$ fluxes higher than $\sim 10^{-17}$ erg s$^{-1}$ cm$^{-2}$ per pixel. Additional tests with a Voronoi binning are being performed to assess whether it improves the results by recovering lower fluxes. Finally, the values of fluxes between both maps cannot be directly compared as no extinction corrections and [NII] removal were applied to the P.E.L.E. map. 

\begin{figure}
\begin{center}
\includegraphics[width=1.0\columnwidth]{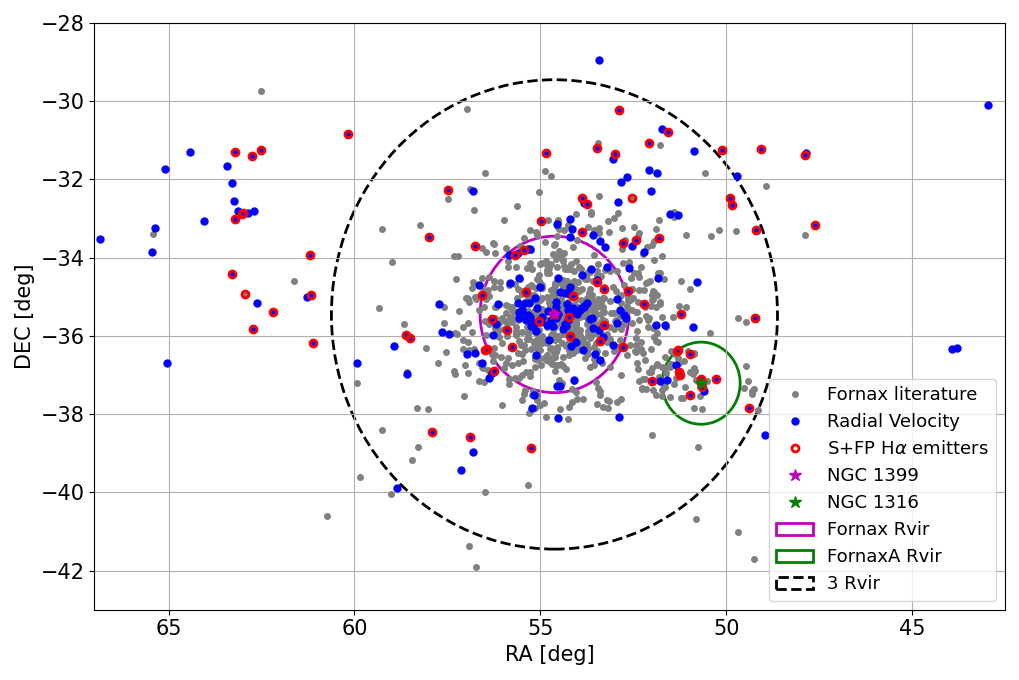}
\caption{Spatial distribution of H$\alpha$+[NII] emitters (red open circles). The 79 emitters are homogeneously distributed within the Fornax virial radius ($R_{vir} = 2$ deg; solid magenta line) and outside $R_{vir}$. Considering only objects with radial velocity information, the fraction of H$\alpha$ emitters inside the Fornax $R_{vir}$ is 15\%, while for the Fornax\,A subgroup ($R_{vir,A} = 1.05$ deg; solid green line) a high fraction of emitters (77\%) is found. The black dashed line represents 3 R$_{vir}$ of Fornax.} 
\label{fig.halpha-dist}
\end{center}
\end{figure}

Using the FLS described in Section\,\ref{subsec:literature}, which consists of 1,005 galaxies, we applied P.E.L.E. to identify H$\alpha$+[NII] emitters. We detected a total of 79 such emitters, all of which exhibit more than five connected pixels with signs of emission. Such a threshold is imposed to eliminate false positives due to residual noise from the images. Figure\,\ref{fig.halpha-dist} presents the spatial distribution of H$\alpha$+[NII] emitters surrounding NGC\,1399 and within the Fornax\,A subgroup, centered in NGC\,1316.

\begin{figure}
\begin{center}
\includegraphics[width=1.0\columnwidth]{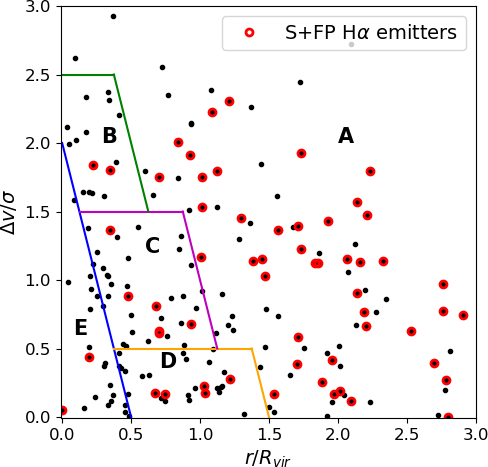}
\caption{Projected Phase Space (PPS) diagram of the H$\alpha$ emitters found using P.E.L.E. results. The solid lines represent the regions defined by \citet{Rhee2017}, where E is dominated by ancient infallers ($t_{\mathrm{infall}} > 8$ Gyr), D contains a mixed population with $t_{\mathrm{infall}}<4-7$ Gyr, and regions A, B and C have mostly galaxies that have fallen at $< 1$ Gyr in the cluster.}
\label{fig.emitters-pps}
\end{center}
\end{figure}

Following \cite{Rhee2017}, we use the Projected Phase Space (PPS) diagram to analyze the infall time ($t_{\mathrm{infall}}$) of the galaxies within the Fornax cluster, considering the separation of the PPS diagram into five regions ranging from ancient infallers ($t_{\mathrm{infall}} < 8$ Gyr) relative to recent ones ($t_{\mathrm{infall}} < 1$ Gyr). Our H$\alpha$ emitters primarily occupy regions associated with recent accretion events, as illustrated in Figure\,\ref{fig.emitters-pps}. Perhaps, the interaction of these recent infallers with the intracluster medium or other cluster galaxies triggered star formation, which would explain the H$\alpha$ emission.

In a forthcoming paper, we will further discuss these H$\alpha$ emitters in relation to other properties, such as galaxy morphology and the spatial distribution of the emission within the galaxy (Lopes et al., in prep.).

\subsection{Density analysis for the Fornax cluster and H$\alpha$ sub-sample}

   \begin{figure}
   \centering
\includegraphics[width=1.0\columnwidth]{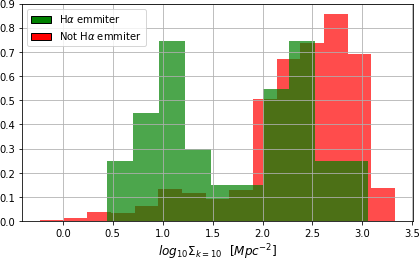}
   \caption{Normalized density histogram for the Fornax IGS. H$\alpha$ emitters are shown in green and non-emitters in red. The H$\alpha$ emitter population shows a distinct second peak at lower densities.}
    \label{img1}
    \end{figure}

We measured the local density of the environment of all galaxies in the Fornax IGS using the $k$-th Nearest Neighbour (KNN) method. Following the rationale described in \citet{SantanaSilva2020} and references therein, we calculated the projected density of each individual galaxy based on the distance to their neighbours, so that smaller distances yield denser environments. To probe cluster-like scales, the $10$th nearest neighbour was chosen ($k=10$). These densities were then compared with the H$\alpha$ emitters data.

   \begin{figure}
   \centering
\includegraphics[width=1.0\columnwidth]{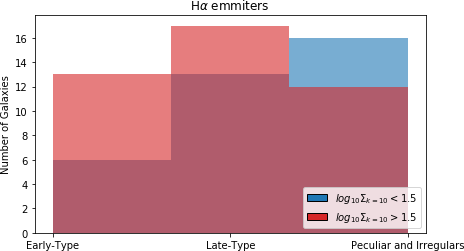}
   \caption{Morphology histogram of the H$\alpha$ emitter sample, divided by density. Emitters in high-density environments (red) show a mixed fraction of early-type, late-type and peculiar and irregular morphologies. On the contrary, low-density emitters are mainly late-type or peculiar or irregular galaxies.}
    \label{img2}
    \end{figure}

Figure\,\ref{img1} shows the normalized histogram of densities for the H$\alpha$ emitters in green and non-emitters in red. The latter sub-sample of galaxies inhabit denser environments with the majority of galaxies at $log_{10}\Sigma_{k=10}/$Mpc$^{-2} \approx 2.8$. However, the H$\alpha$ emitters display a bimodal distribution, with a distinct minimum at $log_{10}\Sigma_{k=10}/$Mpc$^{-2} \approx 1.8$. This distinction is correlated with galaxy morphology as shown in Figure\,\ref{img2}, and as expected according to  the morphology density relation \citep{Dressler1980}: the denser population of H$\alpha$ emitters shows quite similar fractions of early type, late type and peculiar galaxies, while the less dense population is dominated by late type and peculiar and irregular galaxies.

\begin{figure}
   \centering
\includegraphics[width=1.0\columnwidth]{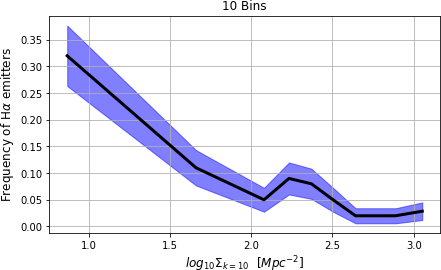}
   \caption{Frequency of H$\alpha$ emitters per bin of density. The black solid line represents the mean values, while the blue shaded region shows uncertainties assuming poissoinian statistics. The dependence of H$\alpha$ emitters as a function of local density is evident.}
    \label{img3}
    \end{figure}

The frequency of H$\alpha$ emitters is a clear function of local density. Figure\,\ref{img3} shows the frequency (number of H$\alpha$ emitters per number of objects in that bin) per density bins. The number of H$\alpha$ emitters is notably higher in low-density regions of the cluster, as expected from previous results \citep[e.g.,][]{Dressler1980,vonderlinden2010,Koyama2011}. This could be evidencing the quenching process exerted by the dense environment.  

\subsection{SED fitting}
\label{subsec:sedfitting}

The 12-band spectral energy distribution (SED) of galaxies produced by S-PLUS can be used to constrain their basic stellar population properties. Methods that make use of photometric data to estimate properties like stellar mass, typical stellar ages, metallicities and dust attenuation (as well as higher-order descriptions of the star formation history of a galaxy) abound in the literature (e.g., \citealt{2021Johnson}; \citealt{BAGPIPES}; \citealt{Delgado_2021}). Although other tools will also be employed in this kind of work within the S+FP, we chose to illustrate results obtained with the \alstar\ code (\citealt{ThainaBatista2023}; \citealt{Delgado_2021}).  \alstar\ performs a non-parametric decomposition of the SED in terms of stellar populations of different ages and metallicities, while simultaneously accounting for the effect of emission lines on the observed photometry. This constitutes a key aspect in the context of S-PLUS because of its narrow bands.

\begin{figure}
    \centering  \includegraphics[width=1.0\columnwidth]{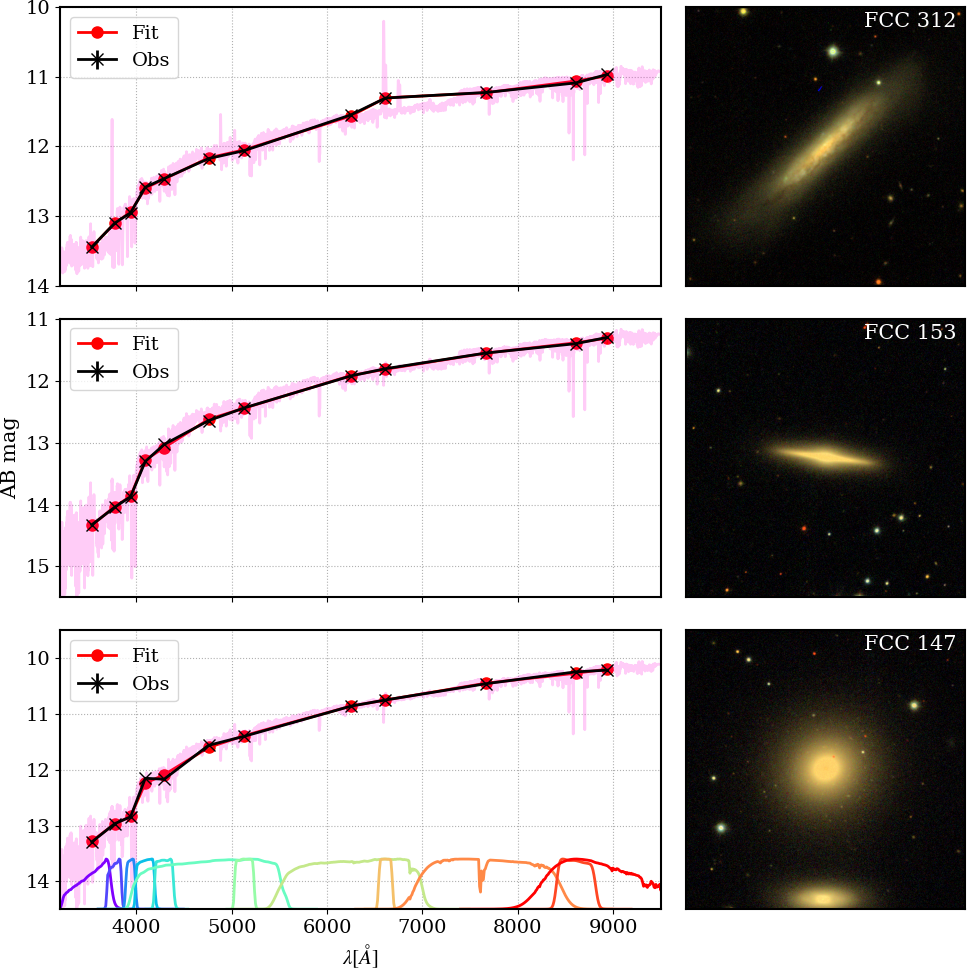}
    \caption{Example of \alstar\, fits of S-PLUS data for three Fornax galaxies: FCC\,312 (top), FCC\,153 (middle) and FCC\,147 (bottom). The data are plotted in black crosses, while the model photometric fluxes are plotted as red circles. The corresponding model spectrum is plotted in magenta.
    }
    \label{fig:AlStar}
\end{figure}

Figure\,\ref{fig:AlStar} illustrates SED fits of three galaxies in the FLS sample (see \citealt{ThainaBatista2023} for details on the ingredients and hypotheses behind these fits). The fits match the observed photometry to within 0.02 mag (or better) on average over the 12 bands. 

The simulations conducted in \citet{ThainaBatista2023} demonstrate that these fits recover stellar population properties of SDSS galaxies derived from a full spectral fitting using similar ingredients by \cite{Werle_2019} to within a good level of agreement. This proves that the amount of information contained in multi-band photometry is not as inferior as that in spectroscopy as one would perhaps expect  \citep{Ferreras2023}.
We have further verified that the inclusion of emission lines in the fits does not lead to any increase of the usual age-dust-metallicity degeneracies characteristic of stellar populations analysis (e.g., \citealp{Conroy2013}).

We determined stellar masses of $\log M_\star/M_\odot = 10.27$, 9.99, and 10.48 for galaxies FCC\,312 (spiral), FCC\,153 (lenticular) and FCC\,147 (elliptical), respectively. These values are within 0.1 dex with those reported by \citet{Iodice2019} and \citet{Raj2019}.
For galaxies FCC\,153 and FCC\,147, we can further compare the mean stellar ages derived from our fits to those reported by \citet{Iodice2019b} based on MUSE-data for the inner 0.5 effective radius.
We obtain mean light- and mass-weighted logarithmic ages of 
$\langle \log t/{\rm yr} \rangle_L = 9.54$ and $\langle \log t/{\rm yr} \rangle_M = 9.97$ for FCC\,153, whereas for FCC\,147 we obtain 
$\langle \log t/{\rm yr} \rangle_L = 10.03$ and $\langle \log t/{\rm yr} \rangle_M = 10.08$.
These values compare well with the $\langle \log t/{\rm yr} \rangle$ values of 10.02 (FCC\,153) and 10.13 (FCC\,147) obtained by \citet{Iodice2019b}.

\subsection{Stellar population parameters with Artificial Intelligence}

Artificial intelligence (AI) methods (machine learning and, in particular, deep learning) offer  alternative approaches to estimating stellar population properties, such as stellar mass and metallicities. These methods establish statistical relationships between photometry and properties. Instead of directly using SED libraries, we introduce a method based on learning from a large and robust data set so that the model is only constrained by the data.  This approach, which involves a known mapping between photometry and stellar population parameters, allows the galaxies' properties to be estimated, even if the emission lines fall out of the narrow bands. This has been done successfully for stellar parameters (e.g. \citealt{Wang_2022}). Here, we extend this procedure to the galaxies in the S-PLUS survey.

Here, we employ algorithms that map S-PLUS photometry to stellar parameters, as those obtained through spectral synthesis by \citet{Werle_2019} (see Section \ref{subsec:sedfitting}), using an extension of the {\sc starlight} code (\citealt{CidFernantesetal2005}) applied to SDSS galaxy spectra. To train machine learning algorithms, we utilized the galaxy catalog from \citet{Werle_2019}. We compute synthetic S-PLUS photometry from their modelled spectra at the Fornax redshift, forming our training dataset (additional details can be found in Cernic et al., in prep.). The training set consists of 137,734 galaxies, each with S-PLUS\textit{-like} photometry and ten stellar population parameters. To solve this regression problem, we tested several machine learning methods, such as Random Forests and XGBoost, but the best one was a Neural Network. A simple architecture of 4 hidden layers was sufficient for the network to obtain each parameter reliably from the photometry. For instance, the stellar mass  exhibited a mean average error of 0.08 dex when compared to the spectroscopic data set. 

\subsection{Low Surface Brightness and Ultra Diffuse Galaxies}
\label{LSBs}

\begin{figure*}
    \centering
    \includegraphics[width=\textwidth]{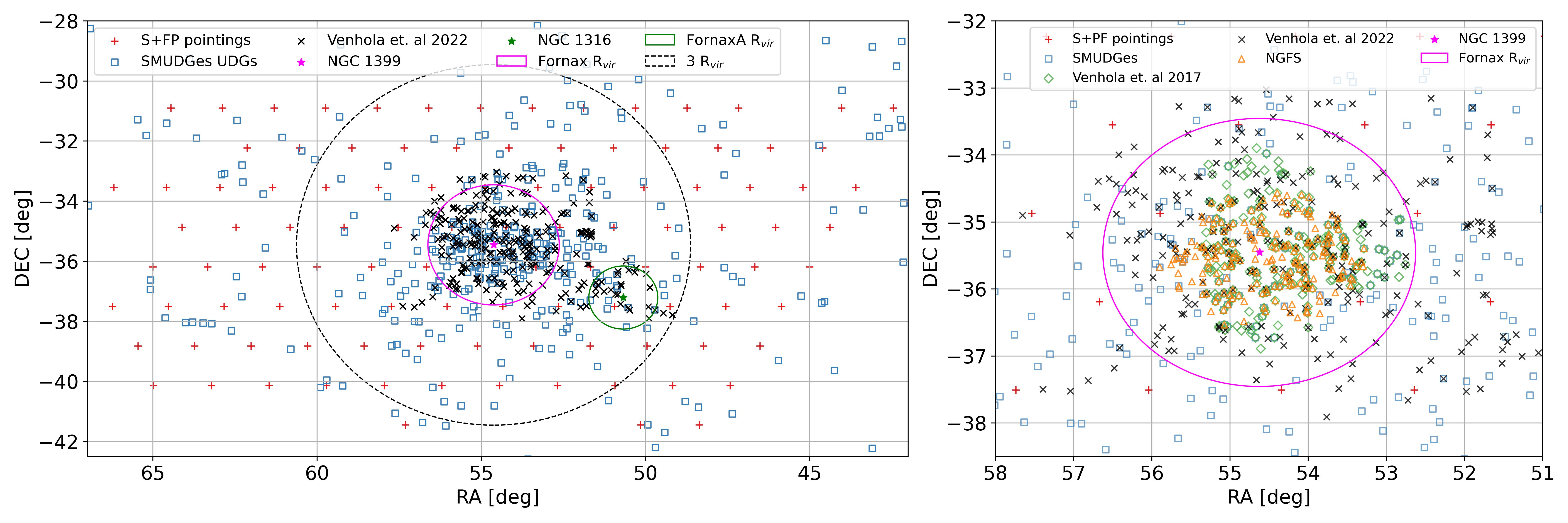}
    \caption{{\it Left:} Projected spatial distribution of the  Ultra Diffuse Galaxies (UDGs) detected in the context of SMUDGes \citep{Zaritsky2023} in the sky area covered by the S+FP. {\it Right:} A zoom into the central region of the Fornax cluster shows the distribution of the SMUDGes subsample compared with the projected distribution of the UDGs reported by the Fornax Deep Survey (FDS; \citealp{Venhola2017,Venhola2022}) and the Next Generation Fornax Survey (NGFS; \citealp{Munioz2015})}.
    \label{Smudges}
\end{figure*}

The population of Low Surface Brightness (LSB) galaxies in the Fornax cluster has been investigated through various surveys, such as the Next Generation Fornax Survey \citep{Munioz2015} and the Fornax Deep Survey \citep{Raj2019}, covering up to approximately three times the R$_{vir}$ and reaching a surface brightness of $\mu \sim$28 mag arcsec$^{-2}$. Ultra diffuse galaxies (UDGs) are a subset of LSB galaxies characterized by their large effective radii ($r_e > 1.5$ kpc). These galaxies have been found in a variety of environments, including the field, groups, and clusters of galaxies. The different properties of UDGs and their globular clusters across environments have raised the question about the role of the environment on UDG evolution and the effects of the environment on their globular cluster (GC) population \citep{Leigh2020}. Recently, \citet{Zaritsky2023} published the complete catalogue of UDGs from the SMUDGes project, which includes about 335 galaxies placed in projection in the 106 S+FP fields (see Figure\,\ref{Smudges}).

As mentioned in Section\,\ref{Intro}, in the context of the S+FP, we are currently studying the galaxy populations in and around the Fornax cluster covering an area of $\sim208$ square degrees. This study extends up to about five times the R$_{vir}$ in right ascension (RA), which is the extension of the FLS (see Figure\,\ref{pointings}). This section introduces our work and presents preliminary results
detecting LSB and UDGs in the Fornax cluster using S-PLUS images. 
The motivation for this search is related to detecting LSB galaxies and UDGs in the (RUN\,1+RUN\,2)$_{res}$ catalogue. 

\begin{figure}
\centering
    \includegraphics[width=0.163\textwidth]{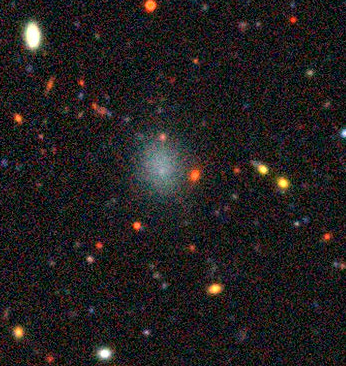}
   \includegraphics[width=0.3095\textwidth]{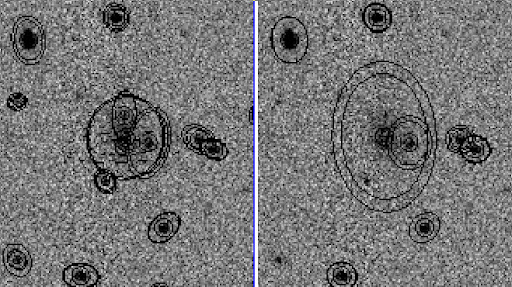}
       \includegraphics[width=0.163\textwidth]{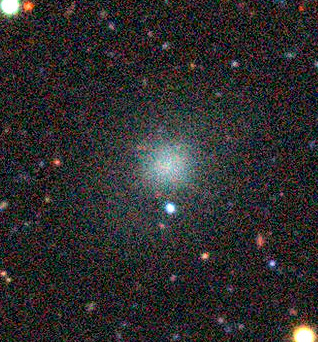}
    \includegraphics[width=0.311\textwidth]{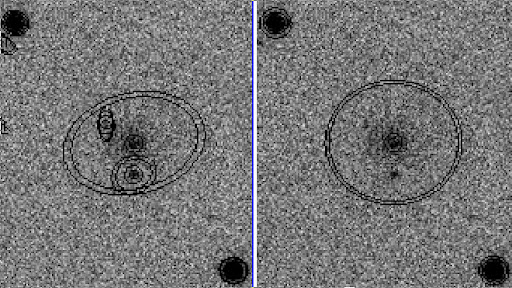}
    \caption{Two examples of LSB galaxies detected by \texttt{SExtractor} in the S+FP images. The left panels correspond to 1.4$\times$1.6 arcmin$^2$ frames of FCC\,271 (top) and FCC\,289 (bottom) obtained from the DESI Legacy Imaging Surveys. The middle and right panels show the g+r+i detection image showing the apertures set by \texttt{SExtractor} on both galaxies using the parameters of RUN\,1 and RUN\,2, respectively. In the case of the detections and measurements made by RUN\,2, which better characterizes the sizes, elongation and position angle of both objects, they present $\mu_\mathrm{max}$($r$-band)=22.8 mag arcsec$^{-2}$ and $\mu_\mathrm{max}$($r$-band)=23.1 mag arcsec$^{-2}$, respectively.}
\label{fig:LSBs}
\end{figure}

Figure\,\ref{fig:LSBs} shows two examples of LSB galaxies in the S-PLUS images detected by \texttt{SExtractor} thanks to the low values adopted for the input parameters DETECT\_THRESH and ANALYSIS\_THRESH in the cases of RUN\,1 and RUN\,2 (see Table\,\ref{tab:Parameters}). The top panels correspond to FCC\,271, and the bottom panels to FCC\,289. The left panels display the images of these two galaxies obtained from the DESI Legacy Imaging Surveys. In contrast, the middle and right panels show the detections made by \texttt{SExtractor} on the S-PLUS images using the parameters of RUN\,1 (middle) and RUN\,2 (right). According to \texttt{SExtractor} RUN\,2 (which provides the best detections for these two objects), FCC\,271 exhibits a maximum surface brightness of $\mu_\mathrm{max}$($r$-band)=22.8 mag arcsec$^{-2}$ while FCC\,289 displays $\mu_\mathrm{max}$($r$-band)=23.1 mag arcsec$^{-2}$. It is noteworthy that, although not visually apparent in the S-PLUS images, \texttt{SExtractor} RUN\,2 seems to detect the faint outer regions of the galaxies based on the sizes of the outermost apertures around the objects. 

\begin{figure}
    \centering
\includegraphics[width=1.0\columnwidth]{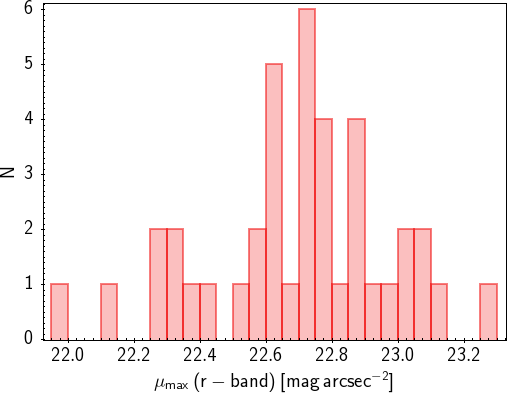}
    \caption{Distribution of the maximum surface brightness ($\mu_\mathrm{max}$) parameter in the $r$-band for the 40 SMUDGes galaxies detected by the S+FP \texttt{SExtractor} runs.}
    \label{histo_UDGs}
\end{figure}

In addition, from a strict (1 arcsec offset) cross-match between the Fornax SMUDGes subsample and the (RUN\,1+RUN\,2)$_{res}$ catalogue, we found 40 objects in common. Figure\,\ref{histo_UDGs} presents a histogram showing the distribution of their $\mu_\mathrm{max}$($r$-band) values. It can be seen that, like in the case of the LSB galaxies detected by our \texttt{SExtractor} runs, these objects display $\mu_\mathrm{max}$($r$-band) in the range 22--23 mag arcsec$^{-2}$. Therefore, from these results, we can provide a surface brightness limit of $\mu_r=23.25$ mag arcsec$^{-2}$ for confident detections in S-PLUS images.

Additionally, we intend to employ a more automated and efficient method based on Convolutional Neural Networks (CNNs), a type of deep learning (DL) architecture. CNNs have proven to be the most successful technique in classifying image data, including galaxy morphology \citep[e.g.,][]{Lukic2019, Zhu2019, Cheng2020}. We plan to train the CNN models using the known LSB galaxies and UDGs in the S+FP fields. By leveraging the power of DL algorithms, we aim to improve the accuracy and reliability of LSB galaxy and UDG detections in the Fornax cluster region and extend the search of these types of objects to other sky regions covered by S-PLUS. This approach can potentially uncover hidden or faint LSB galaxies that previous surveys may have overlooked, contributing to a more comprehensive understanding of the galaxy population in this cluster.

\subsection{Automated Morphological Classification}
\label{subsec:deep-learning}
\begin{figure}
\centering 
\includegraphics[width=0.325\columnwidth]{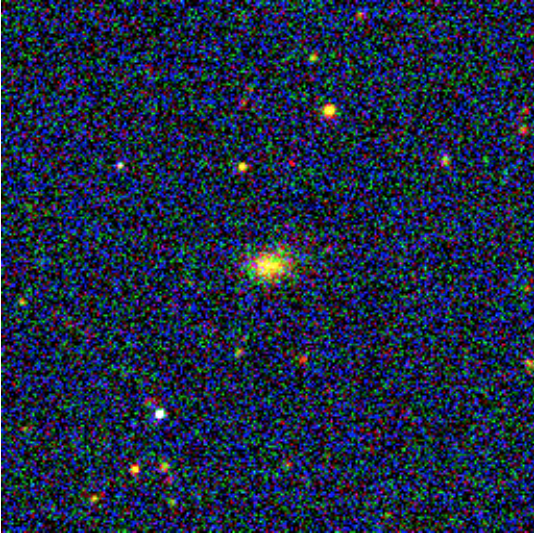}
\includegraphics[width=0.325\columnwidth]{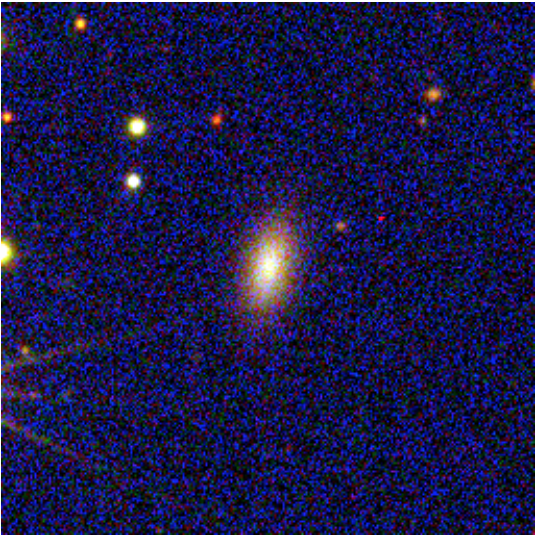}
\includegraphics[width=0.325\columnwidth]{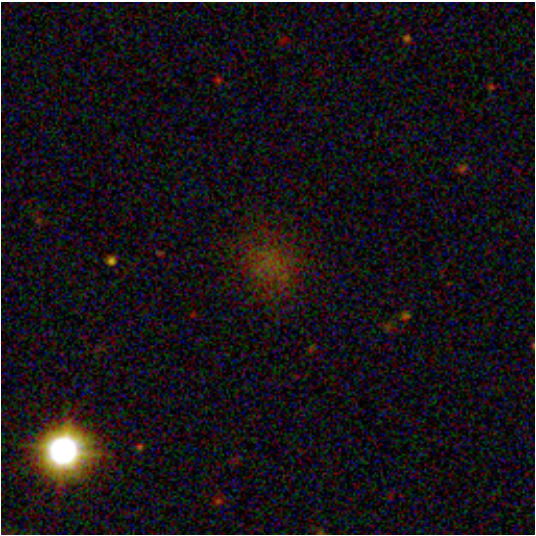}\\
\includegraphics[width=0.325\columnwidth]{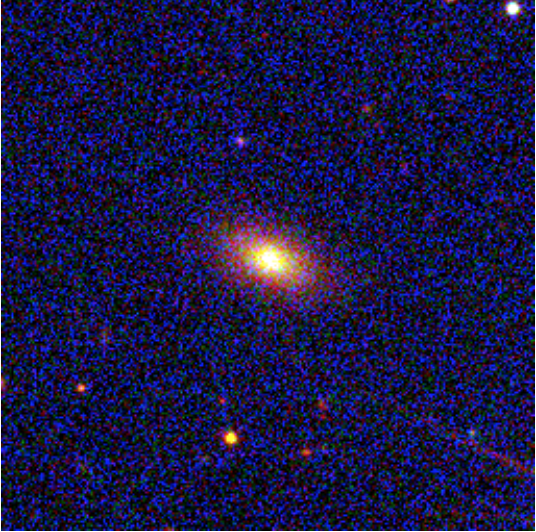}
\includegraphics[width=0.325\columnwidth]{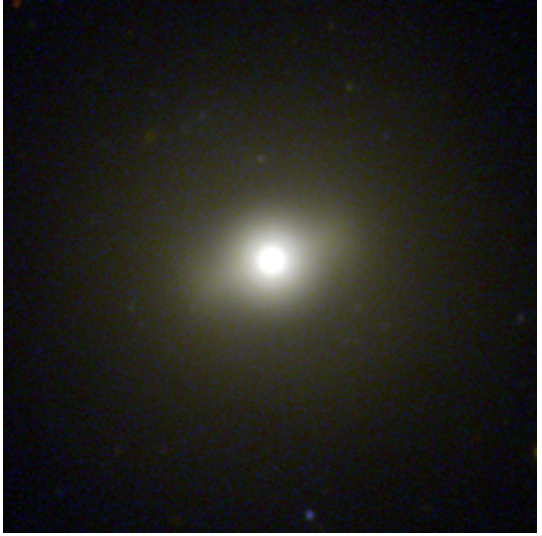}
\includegraphics[width=0.325\columnwidth]{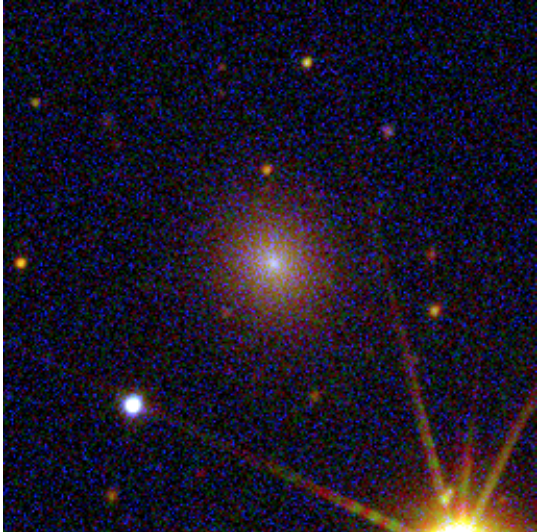}\\
\includegraphics[width=0.325\columnwidth]{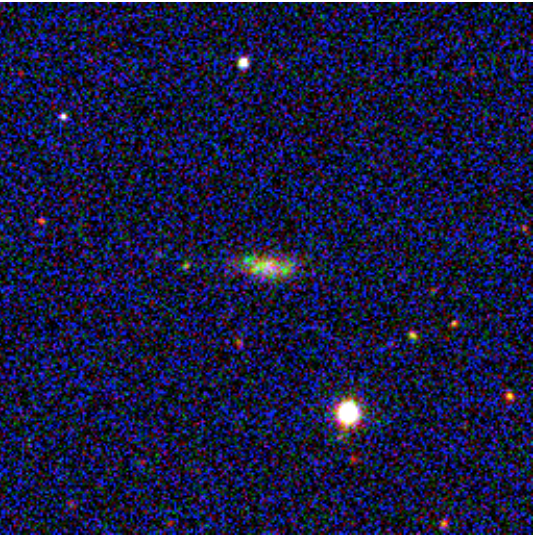}
\includegraphics[width=0.325\columnwidth]{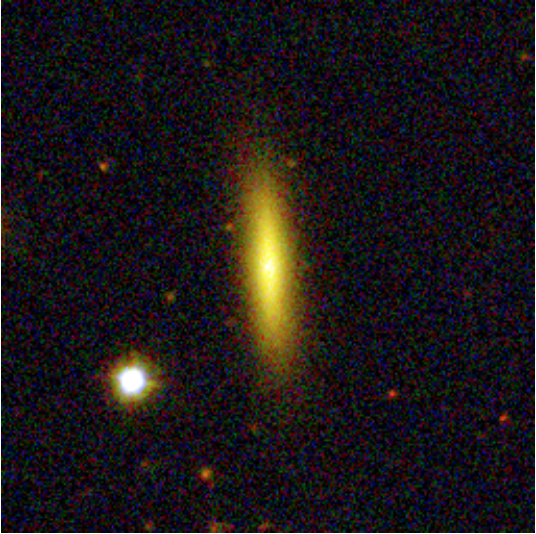}
\includegraphics[width=0.325\columnwidth]{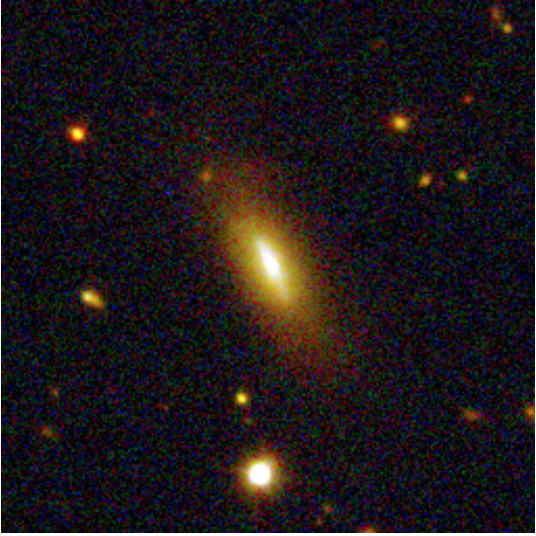}
\caption{Examples of galaxies with a high probability of being spiral but classified as elliptical in the literature ({\it top}), and galaxies classified as S0 in the literature, with a high probability of being early-type galaxies ({\it middle}) or late-type galaxies ({\it bottom}).}
\label{fig:example_morphology}
\end{figure}

Using a DL algorithm from \citeauthor{Bom21} (\citeyear{Bom21,Bom23}), we performed an automated morphological classification of galaxies displaying $r$-band < 18 mag in the S-PLUS DR3, distinguishing between late-type and early-type galaxies. The classification provided information (as a probability of belonging to a specific class) for 156 galaxies confirmed as Fornax members in the literature.  
The automated classification recovers correctly the morphology for all the visually classified late-type galaxies, i.e. spiral and irregular galaxies. More complex is the interpretation of the results for elliptical galaxies, where nearly half of the galaxies classified as elliptical in the literature have a probability higher than 0.5 of being spiral according to the DL algorithm.  We visually inspected all those cases, and we found that they are either lenticular (S0) galaxies (i.e. galaxies with a bulge and a disk), very faint objects, or sources with some hints of star formation (see Figure\,\ref{fig:example_morphology} for three examples).
 
Finally, S0 galaxies should fall in the early-type galaxy group, and most of the galaxies belonging to the S0 galaxy-class in the literature have a probability higher than 0.5 to be early-type galaxies according to the DL algorithm. Yet, as seen for the elliptical galaxies, some of them have a high probability of being spiral galaxies. In Figure\,\ref{fig:example_morphology}, we show some examples of galaxies that in the literature are classified as S0 and have a probability higher than 0.5 of being elliptical (middle) or spiral (bottom) galaxies. The prominence of the bulge and the absence of star formation seem to be the main factors in common between S0 galaxies with a high probability of being early type. At the same time, late-type S0s are disk-dominated and/or star-forming. 

\begin{figure}
\centering
\includegraphics[width=1.0\columnwidth]{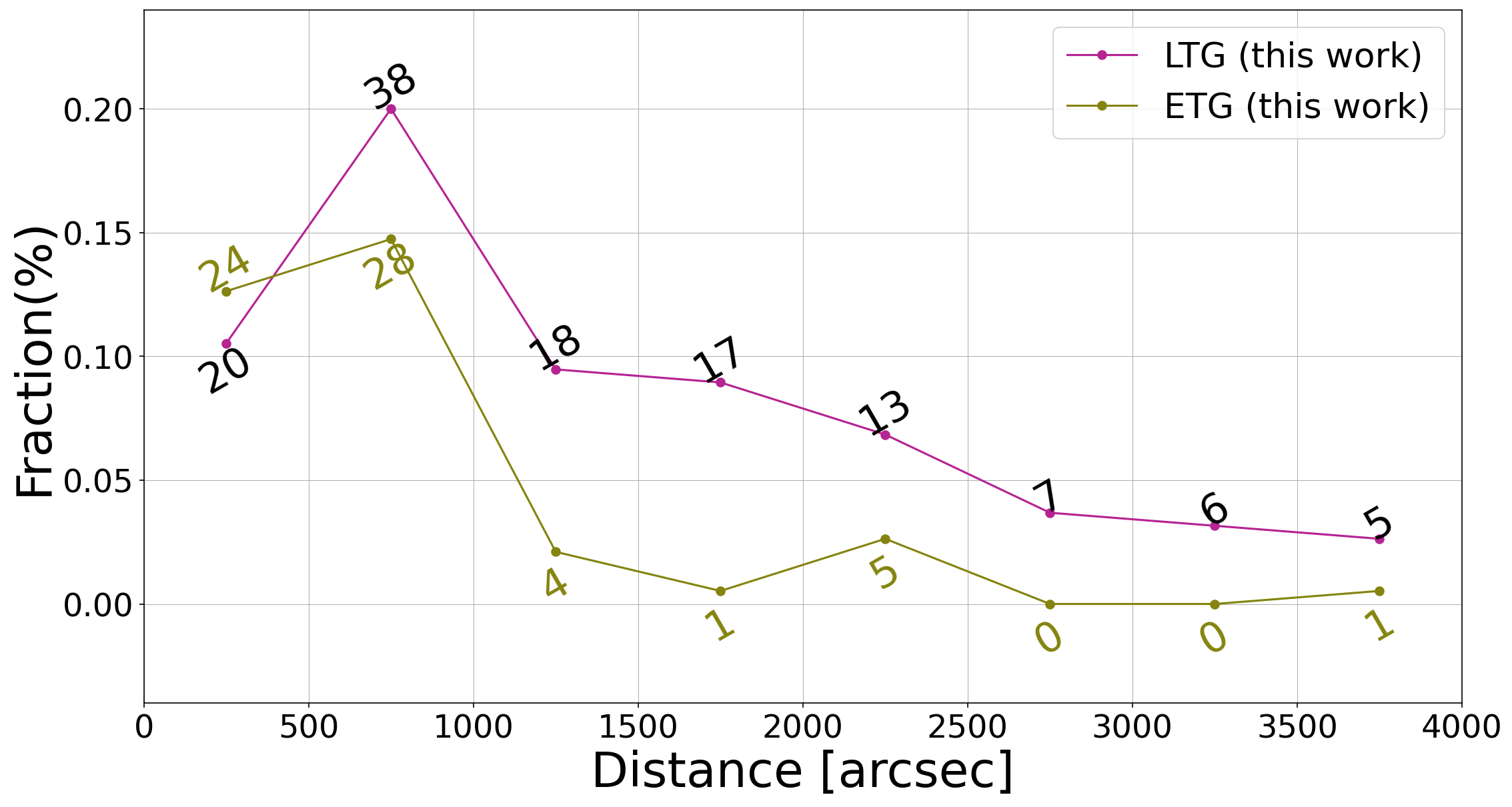}
\includegraphics[width=1.0\columnwidth]{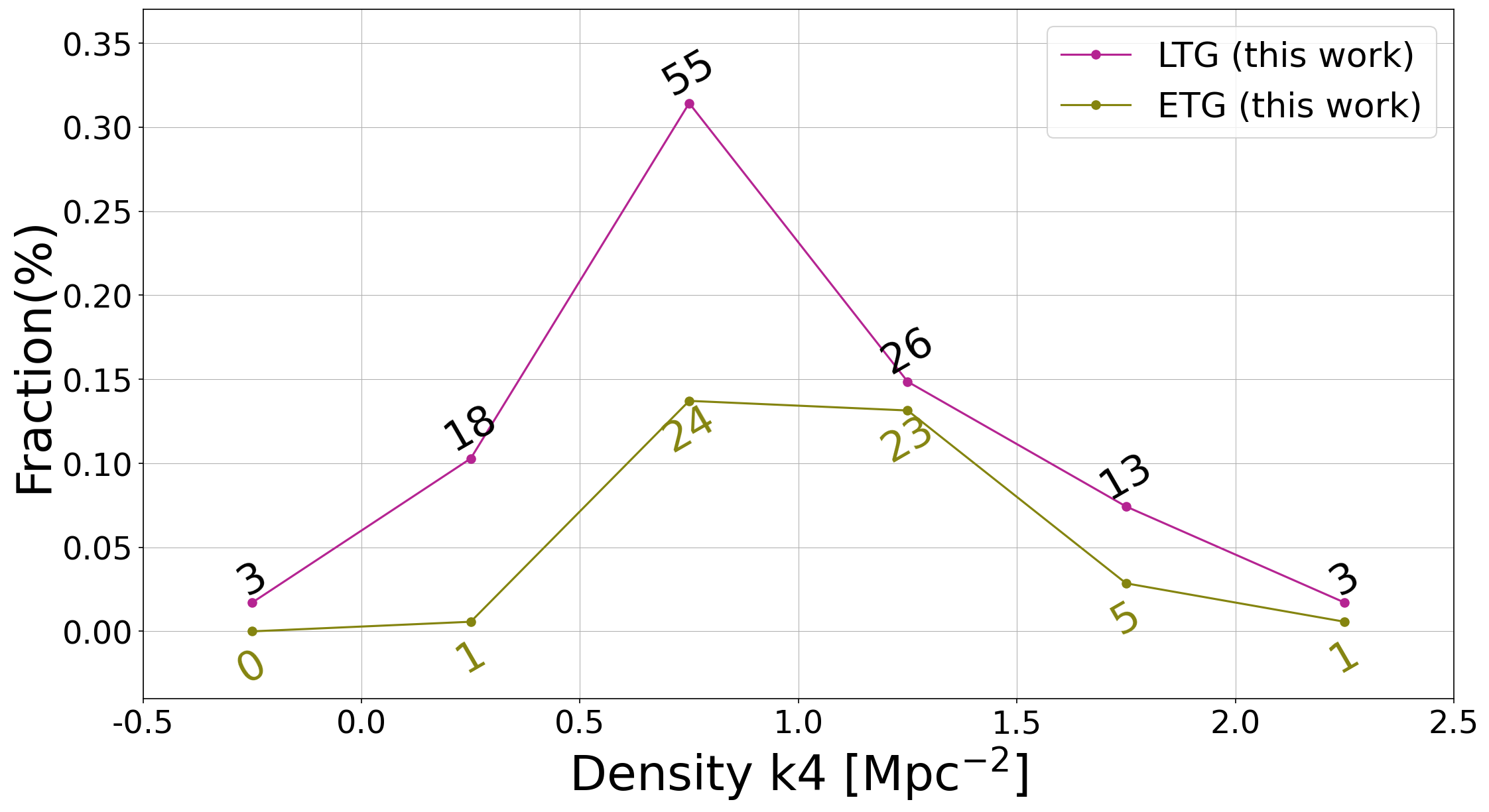}
\caption{Number density of early- (green; ETG) and late-  (magenta; LTG) type galaxies as a function of cluster centric radius {\it(top)}, and of k4 local density {\it(bottom)}.}
\label{fig:dense}
\end{figure}

We employed probability thresholds of 0.68 for late-type and 0.52 for early-type classifications, based on the method described in \citet{Bom23}. This classification yielded 71 late-type and 59 early-type galaxies. We use this same thresholds to calculate the number of early- and late-type galaxies as a function of the distance from the galaxy centre (see the top panel in Figure\,\ref{fig:dense}), as well as of the 4-th nearest neighbour (k4, see \citealt{Bom23} and Silva et al. in prep.) local density (see bottom panel in Figure\,\ref{fig:dense}).

\begin{figure}
\centering
\includegraphics[width=1.0\columnwidth]{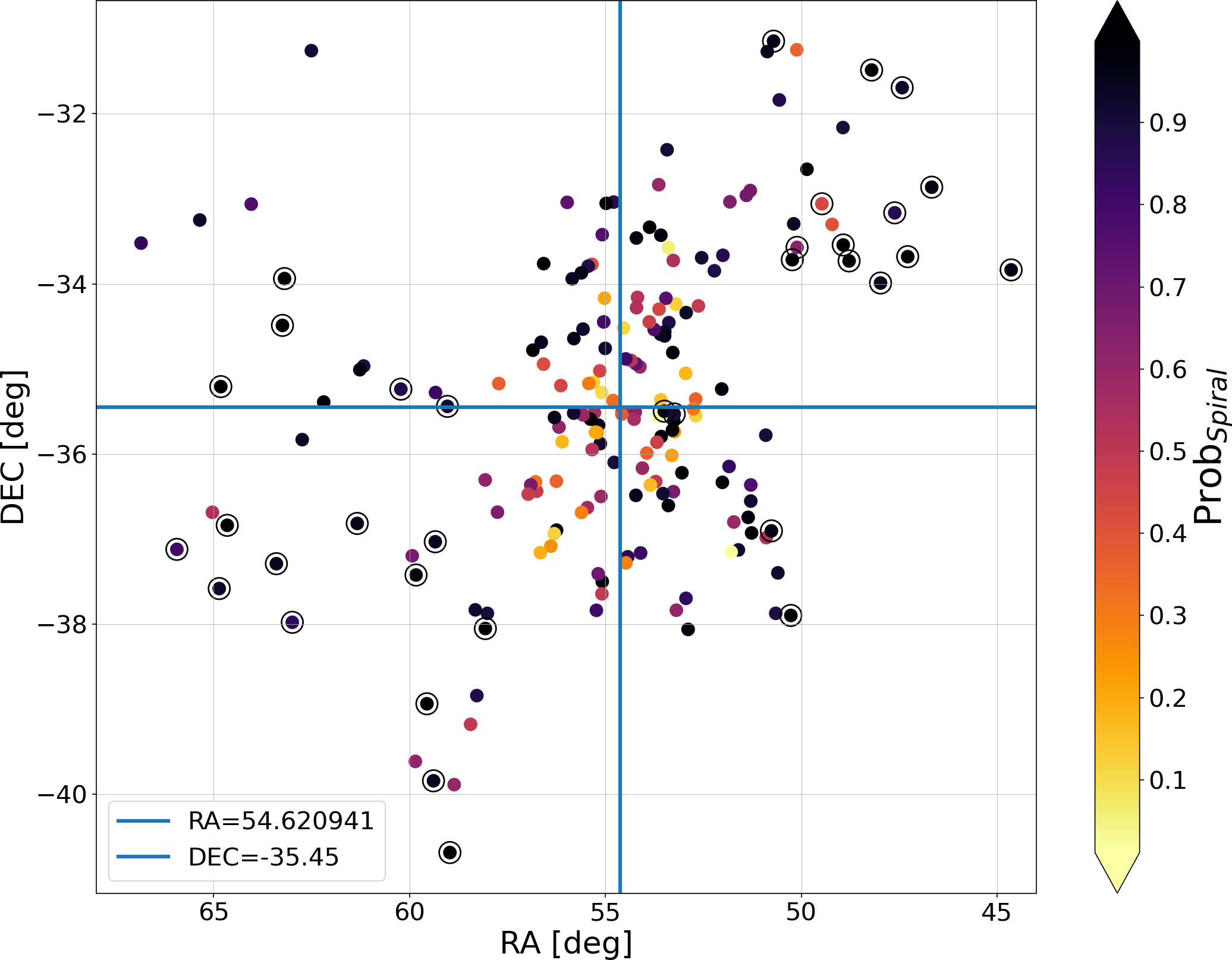}
\caption{Location of spectroscopically confirmed galaxy cluster members (open circles) and new galaxy member candidates from \citet{Bom23} (filled circles without surrounding open circles) colour coded according to the probability of being a spiral galaxy. The RA and DEC coordinates of the vertical and horizontal blue lines depict the location of NGC\,1399.}
\label{fig:distribution}
\end{figure}

The number of early-type galaxies increases toward the cluster centre, as expected, and the dominant morphological type inverts in trend at nearly $R_{vir}$.
Interestingly, the local density, as defined by the 4th nearest neighbor, peaks at a radius corresponding to the location of Fornax\,A.  This indicates that these measurements are sensitive to the presence of substructures.
Moreover, we searched in the catalogue released by \citet{Bom23} for new Fornax member candidates, imposing that the galaxy has a high probability of belonging to the cluster if it has a reliable stamp ProbG\_bin = 1 (see \citealt{Bom23} for more details), a photometric lower redshift (zml; see \citealt{Lima2022}) zml $< 0.02$\footnote{It is not trivial to decide the redshift boundaries of the Fornax cluster using photometric redshifts, given the errors of the zml estimations are of the same order of the cluster redshift. We use the zml errors obtained for spectroscopically confirmed cluster galaxies members to define the maximum redshift at which we can find cluster members.} and odds on the zml estimate higher than 0.4. We visually inspected all the objects to remove duplicated ones. We find 36 new member candidates, out of which 35 are consistent with being late-type galaxies and one an early-type galaxy.
Indeed, all the new candidates are located in the external part of the cluster, as shown in Figure\,\ref{fig:distribution}.
Follow-up studies are necessary to confirm that those candidates actually belong to Fornax, given the error in photometric redshifts at the distance of Fornax. Yet, this simple exercise proves the importance of multi-wavelength surveys, especially combined with accurate DL methods, to identify reliable member candidates, which can further increase the sample of confirmed cluster members.

\subsection{Blue Elliptical Galaxies}
\label{blueE}
A significant fraction of the stellar mass in bright ellipticals seemed to have formed at high redshift. However, in recent decades, many studies have revealed that some elliptical galaxies retain a signature of star formation, most likely residual star formation at late epochs \citep{Zabludoffetal96,Quinteroetal2004,Kavirajetal2007}. The discovery of blue ellipticals \citep{Strateva2001,Fukugitaetal2004,Dhiwar2022} has changed the simplified notion about these galaxies.
\citet{Bamford2009} show that blue elliptical galaxies tend to have $\rm log(M_\star/M_\odot) \le$ 9.6 and reside in low-density environments. \citet{Dhiwar2022} 
find the same trend with environment, even for $L_{*}$ elliptical galaxies. 
On the other hand, \citet{Lacerna2016} find that among the elliptical galaxies in the Coma cluster, about 8\% have blue colours, and $\le 1\%$ are star-forming.

\begin{figure}
\centering
\includegraphics[width=1.0\columnwidth]{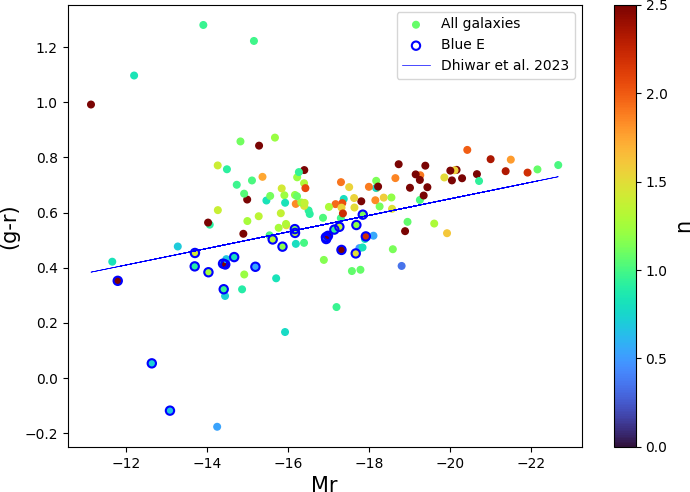}
\caption{Colour-magnitude diagram (CMD) of Fornax cluster galaxies with confident photometry in the 12 S-PLUS filters. Their absolute magnitudes are obtained using standard cosmological parameters and the luminosity distances estimated from the spectroscopic redshifts. The colours are corrected for extinction. Objects are colour-coded according to their S\'ersic index ($n$). Objects below the blue line, identified in \citet{Dhiwar2022} as good separators of the blue cloud from the green valley and red sequence galaxies, are defined as having blue colours. Symbol codes are as in the legend.}
\label{fig:blue_color_mag}
\end{figure}

We explore the existence of blue elliptical galaxies in Fornax using a {colour-magnitude diagram (CMD)}, as shown in Figure\,\ref{fig:blue_color_mag}, where galaxies with blue colours lie below the blue line, which is defined as in \citet{Dhiwar2022}, and galaxies without spiral features are circled in blue. The dots are coloured according to their S\'ersic index ($n$; \citealt{Sersic1968}), as recovered with MFMTK \citep{Ferrari2015}. It is possible to notice that brighter objects tend to have redder colours and higher S\'eric indexes. Most blue ellipticals have $n \simeq 1$, probably including some dwarf spheroidals and LSB galaxies.

\begin{figure}
\centering
\includegraphics[width=1.0\columnwidth]{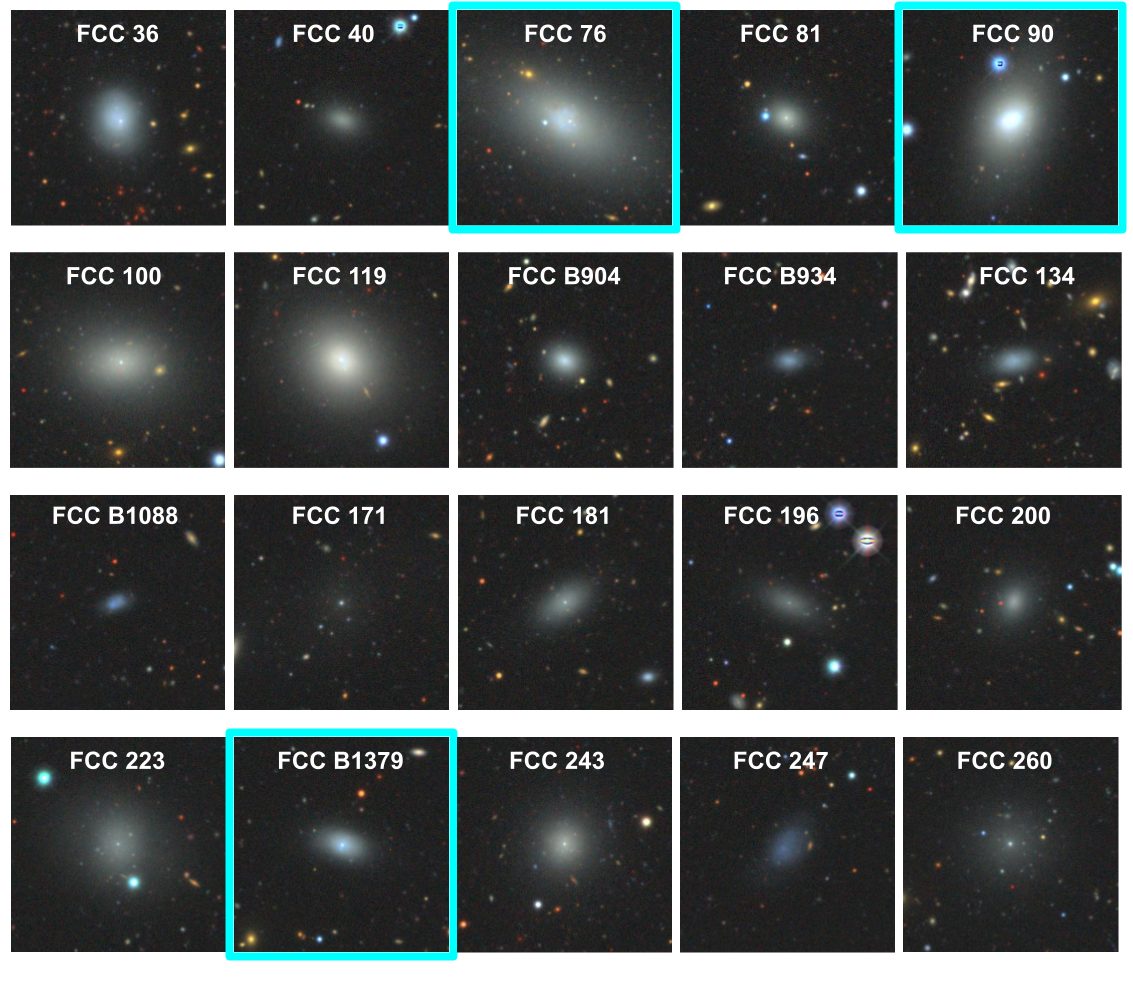}
\caption{Examples of selected blue elliptical galaxies from
 DESI Legacy Imaging Survey. Cyan cells identify galaxies with H$\alpha$ emission.}
\label{fig:blue_snaoshot}
\end{figure}

The images of the selected blue elliptical galaxies from the DESI Legacy Imaging Surveys are displayed in Figure\,\ref{fig:blue_snaoshot}. The depth of these images enables us to identify substructures in the outskirts of these galaxies more effectively.
Cyan squares denote blue elliptical galaxies with H$\alpha$ emission (see Section \ref{subsec:elgs}).
Looking at Figure \ref{fig:blue_snaoshot}, it is clear that the identified elliptical galaxies present different morphologies, i.e. LSB galaxies, dwarf spheroids and massive early-type galaxies, as expected given the S\'ersic index distribution. 
The numbers of blue elliptical galaxies and star-forming blue ellipticals agree with what was found by \citet{Lacerna2016} in the Coma cluster.

\subsection{Globular Clusters}

GCs are among some of the oldest objects in the Universe, making them a key component for understanding the formation and assembly history of galaxies \citep[]{Brodie:2006}. Their relatively high luminosities ($M_V=-5$ to $-10$ mag) and compact sizes (half-light radii of a few parsecs) enable their easy detection in nearby galaxies \citep{Harris:1996}. 
Various GC system properties potentially relevant to cosmological theories of galaxy formation have been identified. These include colour distribution \citep{Larsen:2001, West:2004}, luminosity function \citep{Reed:1994, Whitmore:1995}, radial density distribution \citep{Kartha:2014, Bassino:2006}, specific frequency as a function of galaxy type \citep{Harris:1981, Peng:2008}, total number of GCs as a function of supermassive black hole mass \citep{Burkert:2010, Harris:2011, Harris:2014}, number of GCs as a tracer for the halo mass of the host galaxy \citep{Harris:2013, Burkert:2020} and the nature of their size distribution \citep{Kundu:1998, Larsen:2001, Webb:2012}. The majority of these properties have been exhaustively reviewed in \citet{Brodie:2006}.

One of the most significant efforts in the study of GCs in galaxy clusters is The Next Generation Virgo
Cluster Survey \citep{Ferrarese2012}. 
This survey, conducted as a Large Program with the Canada-France-Hawaii-Telescope using Megacam,   obtained deep imaging in multiple bands (u$^{*}$, g$'$, r$'$, i$'$, z$'$) of the Virgo galaxy cluster. 
The survey comprised 117 pointings covering an area of 104~deg$^{2}$, and a series of papers have already been published using this dataset. One of the main scientific objectives of
this project is the characterization of the faint-end shape of the galaxy luminosity function (g$'$$\sim$25.90 mag). The spatial coverage offered by the S-PLUS survey ($\sim$200~deg$^{2}$) allows us to carry out the largest study of GCs in a galaxy cluster. However, due to its shallower depth, the S-PLUS survey allows us to study only the brightest end of the GC luminosity function (i$'\sim$21.50 mag).

\subsubsection{Globular cluster detection, photometry and selection}

\cite{Buzzo:2022} assessed the effectiveness of identifying GCs around NGC\,1399 using S-PLUS images. Despite the shallow depth of S-PLUS, that limited their work to $r<21.3$ mag, they measured reliable photometry and performed robust SED fitting for a sample of 115 GCs (85\% completeness), recovering ages, and metallicities, as well as photometric redshifts consistent with that of NGC 1399. For a more detailed explanation on how they selected the GC candidates and about their SED and template fitting techniques, we refer the reader to \citet{Buzzo:2022}. For their photometric redshift estimates (and, consequently, radial velocity estimates), they recovered an accuracy of 0.001 down to $r<17$ mag, 0.002 down to $r<19$ mag and 0.003 down to $r<21.3$ mag (S-PLUS DR3 $3\sigma$ limit) for the entire S-PLUS DR3. These statistics, however, are dominated by galaxies beyond the local Universe. For closeby sources, the uncertainties are higher such that at the distance of the Fornax cluster, uncertainties are of the same order as the photo-z itself. Even with the high uncertainties on the photo-zs, \citet{Buzzo:2022} were able to show that it is possible to recover ensemble properties of GC systems at the distance of the Fornax cluster and, therefore, can be used to find/derive properties of GCs in this work.

Lomel\'i-N\'u\~nez et al. (in prep.) continued the photometric study of GCs in Fornax using the 106 pointings of the S+FP, analysing the 12 bands in the $\sim$23$\times$11~deg$^2$ FoV observed with S-PLUS. 
They utilized the \texttt{SExtractor} software  \citep{Bertin1996} for source detection. 
The S-PLUS images have low resolution (0.55~arcsec~pixel$^{-1}$) and are shallower ($\sim$100-800~s for blue and red bands, respectively) in comparison, for example, to those used by  \citet{Iodice2016} (0.21~arcsec~pixel$^{-1}$) and \citet{Cantiello:2018} (0.26~arcsec~pixel$^{-1}$). However, our spatial coverage is wider than the analyzed in those works as they cover only the central region of our pointings. 
When we compare our detections with the catalogue of spectroscopically confirmed GCs of \citet{Avinash:2022}, we can recover $\sim$1000 out of a total of 2268 sources.

The PSF measurements were conducted using a combination of \texttt{SExtractor} and \texttt{PSFEx}\footnote{\url{https://www.astromatic.net/software/psfex/}} \citep{Psfex:2011}. The PSF photometry methodology is comprehensively outlined in \citet{Lomeli:2017}, \citet{Lomeli:2022} and \citet{Lomeli:2022cfht}. It consists of the following steps: 

\begin{itemize}
    \item a first run of \texttt{SExtractor} to detect and select sources for the PSF creation; 
    \item the creation of the PSF with the \texttt{PSFEx} software using the point sources selected in the previous step; 
    \item a second run of \texttt{SExtractor} using the PSF created in the previous step to obtain the PSF magnitude (MAG\_PSF). 
\end{itemize}

\begin{figure*}
    \centering
    \includegraphics[width=1.0\columnwidth]{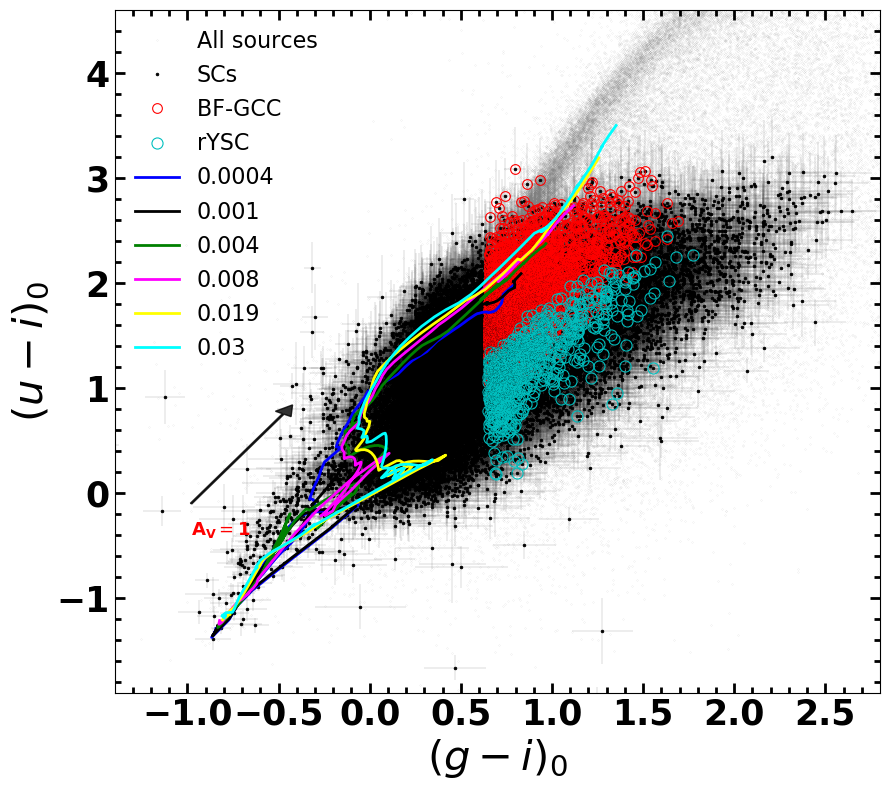}
    \includegraphics[width=1.0\columnwidth]{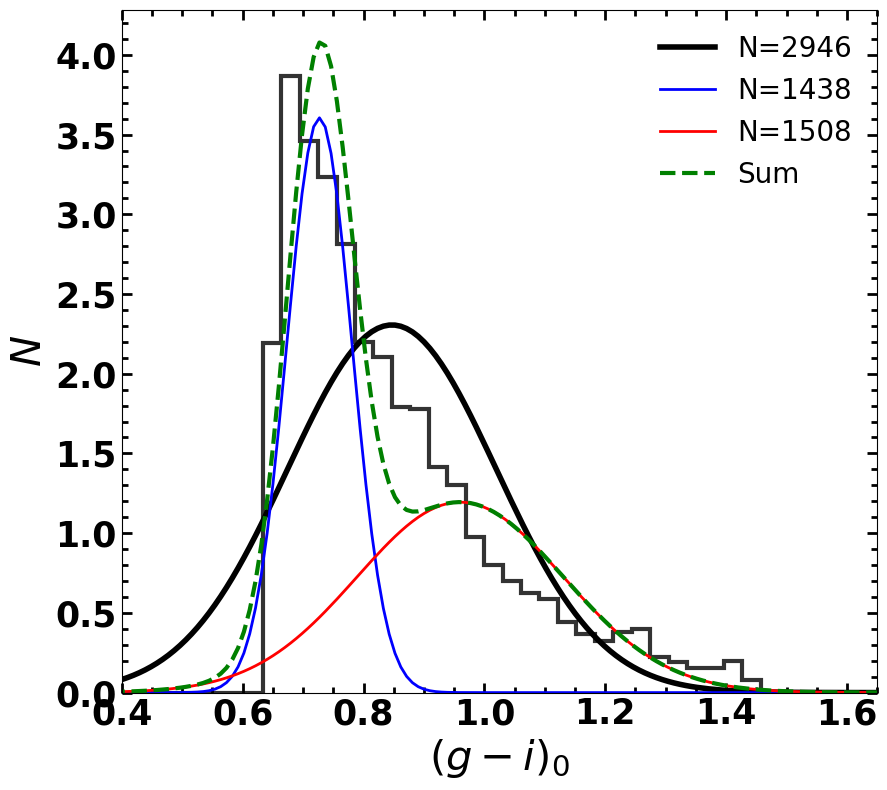}
    \caption{Main properties of the globular cluster (GC) candidates in the Fornax cluster. 
     {\it Left panel:} colour-colour diagram.    
     The $u$-selected GC candidates in the $(u-i)_0$ vs $(g-i)_{0}$ diagram. Bona-fide GC candidates (BF-GCC) with $(g-i)_{0}>$0.65~mag are shown as red dots.  
     The evolutionary loci of Single Stellar Populations (SSPs) from \citet{Bruzual2003} for different metallicities, using a Kroupa initial mass function, are shown by the solid curves of different colours, following the colour notation shown in each panel. 
     The reddening vector corresponding to $A_{V}=1$~mag is also shown. 
     The reddened young stellar clusters (rYSCs) that occupy the GC colour-colour region are contaminants, and are identified by black dots surrounded by cyan circles. 
     Sources that match GC structural parameters, possible star clusters (SCs), are shown as black dots.
     The cloud of small gray dots are all sources detected.
     {\it Right panel:} colour distribution for BF-GCC. Different fits are shown:
     an unimodal distribution for the entire distribution ({\it black solid} line) and a bimodal distribution for the bluest part ($\mu=0.72$) of the distribution ({\it blue solid line}) and for the reddest part ($\mu=0.95$) of the distribution (red solid line). The {\it green dashed line} is the sum of the red and blue lines.   
    } 
    \label{fig:gcs_results} 
\end{figure*}

Different methods for the GC candidate selection can be found in the literature. Lomelí-Nu\~nez et al. (in prep.) used a combination of \texttt{SExtractor} structural parameters, \fwhm, \class, \fluxradius\  and \spread, to define a cluster sample, and a photometric parameter, colour, to separate young stellar clusters from GCs. 
Finally, they used the redshifts estimated via  template fitting using  LEPHARE\footnote{\url{http://www.cfht.hawaii.edu/~arnouts/ lephare.html}} \citep[][]{Arnouts:2002}  to reject galactic or extragalactic outliers. The \citet{Avinash:2022} catalogue was used to delimit the structural parameters selection criteria, choosing for that the $i$-band structural parameters. Additionally, all sources with a {\sc flag}$\neq$0\footnote{The {\sc flag} corresponds to a warning about the source extraction process. Different values of {\sc flag} indicate various problems with the photometry. For instance, sources with close bright neighbors are assigned a {\sc flag}=1; blended objects, {\sc flag}=2; saturated sources, {\sc flag}=4. A combination of these values is also possible. See the \texttt{SExtractor} manual for further explanation.} in $i$-band are rejected. The {\sc flag}=0 parameter excludes sources that present problems in \texttt{SExtractor} detection, e.g., sources with close bright neighbors, blended objects and saturated sources. Also, all the sources have an error in the PSF photometry, {\sc magerr\_psf}$_{i-band}\leq$0.2.

To establish the magnitude at which the sample is 50\% complete, a series of completeness simulations were conducted specifically in the $i$-band. The magnitude mean value at which the sample is 50\% complete, m$_{i,50}\sim21.50\pm$0.1~mag, is obtained from the 106 S+FP pointings. A restriction was made in the brightest part of the $i$-band GC Luminosity Function (GCLF) (m$_{i}$>=19.5~mag) to avoid other contaminants (e.g., Galactic stars and ultra compact dwarfs). All the data were corrected for Galactic extinction with the \cite{Schlafly:2011} values given for S-PLUS iDR4: $A_{g}$=0.0414 and $A_{i}$=0.0204.

GCs' colour is the most useful discriminator between young and old populations. For example,
a metal-poor    (Z$\leq$0.001) simple stellar population (SSP) predicts $(g-i)\geq$ 0.65~mag for populations older than $\sim$3~Gyr \citep{Bruzual2003}. In the left panel of Figure\,\ref{fig:gcs_results}, we show the $(u-i)_{0}$ vs $(g-i)_{0}$ colour-colour diagram, which has been corrected for galactic extinction. This diagram showcases the sources that were selected using both structural parameters and the magnitude cuts. The use of colour–colour diagrams involving ultraviolet and optical filters is known to break the age-reddening degeneracy (e.g., \citealt{Georgiev:2006}; \citealt{Bastian:2011}; \citealt{Fedotov:2011}; \citealt{Lomeli:2022}; \citealt{Whitmore:2023}). Finally, after filtering using the above structural, magnitude and colour criteria, it was possible to obtain a catalogue of 28,713 GC candidates from $\sim 3 \times 10^6$ sources. A final filtering of the GC candidates was made by comparing with the Gaia Data Release 3 (Gaia DR3;  \citealt{GAIADR3:2021}) catalogue centered on NGC~1399 with a coverage of 10~deg$^{2}$. The number of sources from the Gaia catalogue is 1,129,284. However, for the comparison we only used the sources with a proper motion $S/N_{\mu}<3$\footnote{The non-stars are expected to have proper motions that are consistent with 0 at the $3 \sigma$ confidence level, while genuine stars are expected to have $S/N_{\mu}$ > 3.} \citep{Voggel:2020, Buzzo:2022}. The final catalog was assembled using a methodology similar to that of \citet{Lomeli:2022} and \citet{Buzzo:2022}. This catalog comprised $\sim1\times10^{4}$ objects, categorized into three groups: 
\begin{itemize}
\item[a)] {\it GC candidates:} black dots with $(g-i)_{0}>$0.65~mag in the left panel of Figure~\ref{fig:gcs_results};
\item[b)] {\it Bona-Fide GC candidates (BF-GCC):} red dots in the left panel of Figure~\ref{fig:gcs_results};
\item[c)] {\it reddened Young Stellar Clusters (rYSCs):} 
cyan dots in the left panel of Figure~\ref{fig:gcs_results}. 
\end{itemize}
Here we present a brief analysis of the results in $(g-i)_{0}$ colour only for the BF-GCC sample.

\subsubsection{First result on the Fornax GC system}

In the right panel of Figure\,\ref{fig:gcs_results}, we show the $(u-i)_{0}$ versus $(g-i)_{0}$ colour-colour distribution for the BF-GCC. In the plot, it can be seen that it is clearly not well fit by a single non-skewed Gaussian, and in common with other GC colour distributions likely represents a bimodal distribution \citep[e.g.,][]{Larsen:2001, Brodie:2006, Cantiello:2007}.
 
In different studies, it has been shown that a bimodal distribution in colour can be interpreted as a bimodality in metallicity \citep[e.g.,][]{Brodie:2006}. We used the Gaussian Mixture Modeling (GMM) code \citep{Muratov:2010} to fit a bimodal distribution. For a unimodal distribution (black solid line), the GMM fit yielded parameters $\mu=0.84\pm0.01$ and $\sigma=0.17\pm0.01$. In contrast, for a bimodal distribution, the fitting parameters for the bluest part of the distribution (blue solid line) were $\mu=0.72\pm0.01$ and $\sigma=0.05\pm0.01$, while for the reddest part of the distribution (red solid line), the parameters were $\mu=0.95\pm0.01$ and $\sigma=0.17\pm0.01$. For a distribution to be considered as bimodal, the Kurtosis must be negative, the distance between the peaks of the distributions (D) must be greater than 2 and $p$-values (p$_{\chi^{2}}$=0.01, p$_{DD}$=0.15, p$_{kurt}$=0.2) must be small. We found a Kurtosis of -1.07 and a separation D$=1.82\pm0.03$. From the GMM statistical results for a double Gaussian fit, we obtain that the distribution is nearly (6$\sigma$ in D) a bimodal distribution.

In Lomel\'i-N\'u\~nez et al. (in prep.), a detailed study of the properties of the Fornax GC system in the 12 S-PLUS bands is carried out. That study will include colour distributions, luminosity function, spatial distribution, number of GCs, and specific frequency ages and metallicities using the SED fitting method. 

\subsection{Fornax-like systems in cosmological simulations}

Given the complex interplay of physical processes affecting the evolution of galaxies, in general (e.g., star formation, chemical enrichment, supernova feedback, etc), and in dense environments, in particular (e.g. ram pressure stripping, starvation), cosmological hydrodynamical simulations constitute crucial tools to constrain plausible formation scenarios for observed galaxy populations at low redshift.
As part of the S+FP, we plan to compare properties of observed galaxies in Fornax with those predicted for similar systems by state-of-the-art numerical simulations. On the one hand, this analysis will help to determine to what extent current simulations can reproduce the observed complex features in clusters such as Fornax. On the other hand, this study will contribute to explain the origin of observed properties in Fornax by constraining the plausible formation scenarios of similar simulated clusters. 

An analysis of Fornax-like galaxy clusters in the {\sc{eagle}} (e.g. \citealp{Schaye2015}, \citealp{Crain2015}) and {\sc{IllustrisTNG}} (e.g. \citealp{Springel2018}, \citealp{Pillepich2018}) cosmological numerical simulations is being carried out, selecting them according to observed properties. In both suites of simulations, all the relevant physical processes that drive galaxy formation and evolution (e.g. star formation, chemical enrichment, gas heating, radiative cooling, feedback from supernovae, and feedback from active galactic nuclei, among other processes) are taken into account following the joint evolution and merger histories of dark matter and baryonic structures. A $\Lambda$CDM cosmology is adopted, with cosmological parameters taken from \citet{Planck2016}: $h = 0.677$, $\Omega_\Lambda=0.693$, $\Omega_m=0.307$, $\Omega_b=0.04825$, and $Y = 0.248$. Complete details of each simulation and the implemented physics can be found in the corresponding {\sc{eagle}} and {\sc{IllustrisTNG}} data releases (e.g. \citealp{McAlpine2016} and \citealp{Nelson2019}, respectively). From the {\sc{eagle}} suite, we are studying simulated galaxy clusters extracted from the reference model, labelled as RefL0100N1504, while from the  {\sc{IllustrisTNG}} simulations, the TNG100-1 and TNG300-1 models are being used. The {\sc{eagle}} RefL0100N1504 and {\sc{IllustrisTNG}} TNG100-1 simulations have a simulated box size of $L \approx 100$ comoving Mpc, and a baryonic and dark matter mass resolution of $m_{\rm b} \approx 1.6\times10^6~{\rm{M}}_\odot$ and $m_{\rm DM}\approx 8.5\times10^6~{\rm{M}}_\odot$. On the other hand, the {\sc{IllustrisTNG}} TNG300-1 simulation has a larger box size ($L \approx 300$ comoving Mpc) but a lower mass resolution ($m_{\rm b}\approx1.1 \times 10^7~{\rm{M}}_\odot$ and $m_{\rm DM}\approx 6\times10^7~{\rm{M}}_\odot$).

\begin{figure*}
	\centering
    \includegraphics[width=\textwidth]{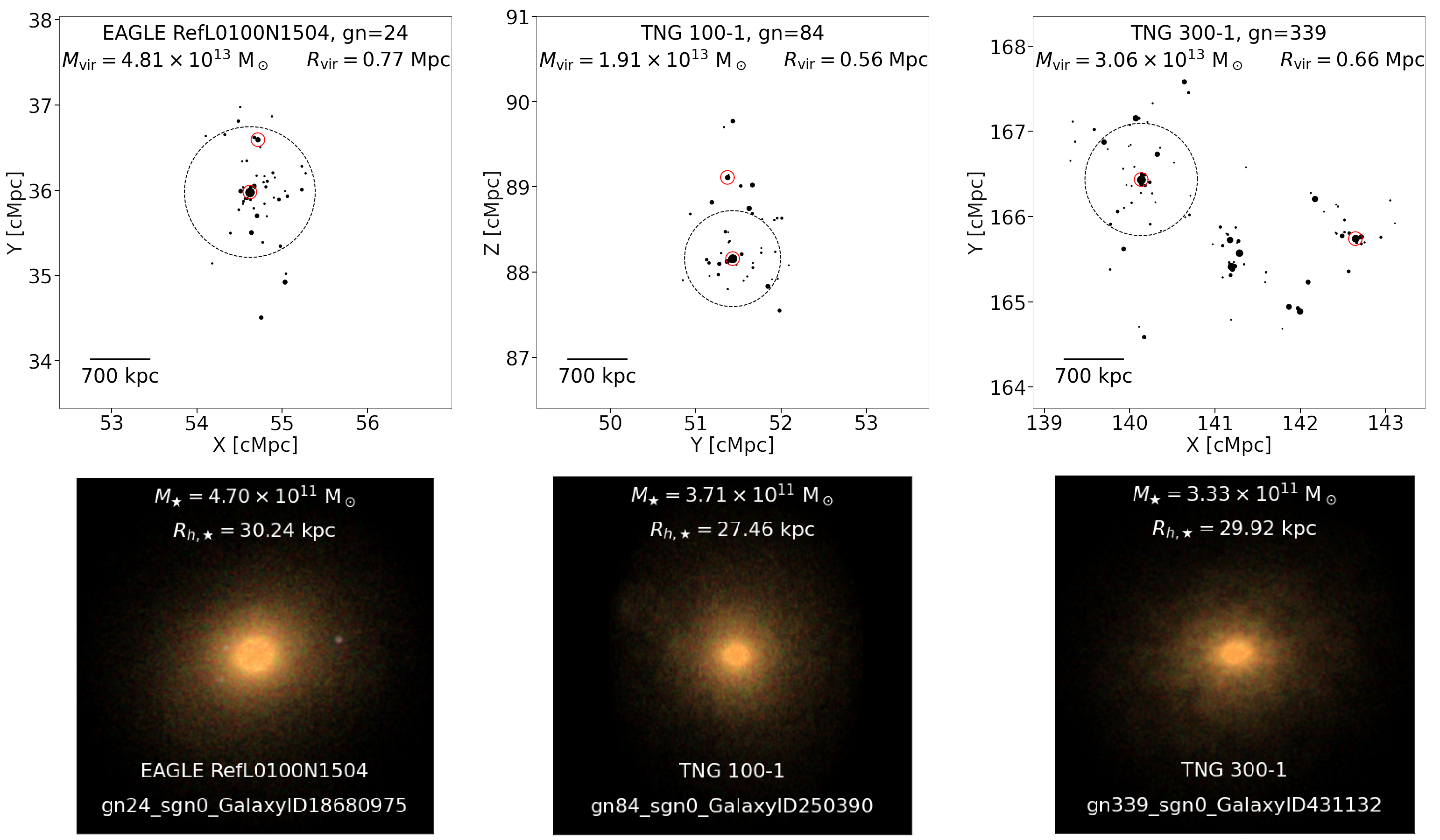}
	\caption{{\textit{Top}}: schematic view of three simulated Fornax-like clusters, extracted from {\sc{eagle}} RefL0100N1504 (left), {\sc{illustris-tng}} TNG100-1 (middle) and TNG300-1 (right). Each symbol represents a galaxy of the simulated cluster, with its size denoting the stellar mass of the galaxy. Dashed black circles enclose the cluster virial radius, centred on the central galaxy. Solid red circles enclose the central galaxy and the second most massive galaxy in the cluster. The cluster `GroupNumber', virial mass and virial radius are listed in each panel. {\textit{Bottom}}: mock images of the central galaxies of simulated Fornax-like clusters shown in the top panels. The stellar mass $M_\star$, stellar half-mass radius $R_{{\rm h,}\star}$, and galaxy identifier are listed. Images were generated with the composition of the $gri$ S-PLUS bands using the scheme of \citet{Lupton2004}.
	}
    \label{fig:fig1_simus}
\end{figure*}

\subsubsection{Sample of simulated Fornax-like clusters}

From the aforementioned numerical simulations, we selected clusters{\footnote{More precisely, in the simulations, a `cluster' of galaxies is a large-scale structure identified with the friends-of-friends algorithm (FoF; \citealt{Davis1985}), and they are commonly referred to as `groups' or `halos'.}} at redshift $z = 0$ with a virial mass{\footnote{In simulations, we define the virial mass, $M_{\rm 200}$, of a simulated cluster as the total mass within $R_{\rm 200}$, which is the physical radius within which the mean internal density of the cluster halo is 200 times the critical density of the Universe. The corresponding halo is centred on the cluster dark matter particle with the minimum gravitational potential.}} of $10^{13}~{\rm{M}}_\odot \leqslant M_{\rm vir} \leqslant 10^{14}~{\rm{M}}_\odot$ (considering that the estimated virial mass of the Fornax cluster is $M_{\rm vir, Fornax} \approx 6 \times 10^{13}~{\rm{M}}_\odot$; \citealt{Maddox2019}). In order to refine our selected sample, and to find simulated clusters that closely resemble Fornax, we selected clusters in which the central galaxy is similar to NGC\,1399 (the central galaxy in Fornax) in terms of mass, size and morphology. Taking into account that NGC\,1399 has a stellar mass $M_{\star,\rm{NGC\,1399}} \approx 2.8 \times 10^{11}~{\rm{M}}_\odot$, a half-light radius $R_{\rm NGC\,1399} \approx 30~{\rm kpc}$, and a visually spherical morphology (see e.g. \citealt{Iodice2019}), we restrict our sample to clusters with a central galaxy having a stellar mass $1 \times 10^{11}~{\rm{M}}_\odot \leqslant M_\star \leqslant 5 \times 10^{11}~{\rm{M}}_\odot$, a half-mass radius{\footnote{Half-mass radii of simulated galaxies are directly available in the public databases, not so the half-light radii. These quantities are not exactly equal, but their values should be similar.}} $20~{\rm kpc} \leqslant R_{\rm{h},\star} \leqslant 40~\rm{kpc}$, and a visually spherical morphology. With our selection criteria, we obtained 10 simulated Fornax-like clusters from {\sc{eagle}} RefL0100N1504, 6 clusters from {\sc{IllustrisTNG}} TNG100-1, and 29 clusters from TNG300-1. In order to avoid resolution issues, we have chosen to consider only simulated galaxies within each selected cluster with $M_\star \geqslant 10^8~{\rm{M}}_\odot$. This lower stellar mass limit corresponds to $\sim 100$ star particles in {\sc{eagle}} RefL0100N1504 and {\sc{IllustrisTNG}} TNG100-1, and $\sim 50$ star particles in TNG300-1. For a given simulation, each simulated cluster is identified with an unique number (`gn' or `GroupNumber') within that simulation. In addition to this, each galaxy is completely identified with the `GroupNumber' of the cluster it belongs to, and two additional numbers (`sgn' or `SubGroupNumber', and `GalaxyID'). The `GalaxyID' identifier is unique for each galaxy within the simulation, while `sgn' can be used to identify central (sgn=0) and satellite (sgn>0) galaxies.

\begin{figure*}
	\centering
    \includegraphics[width=\textwidth]{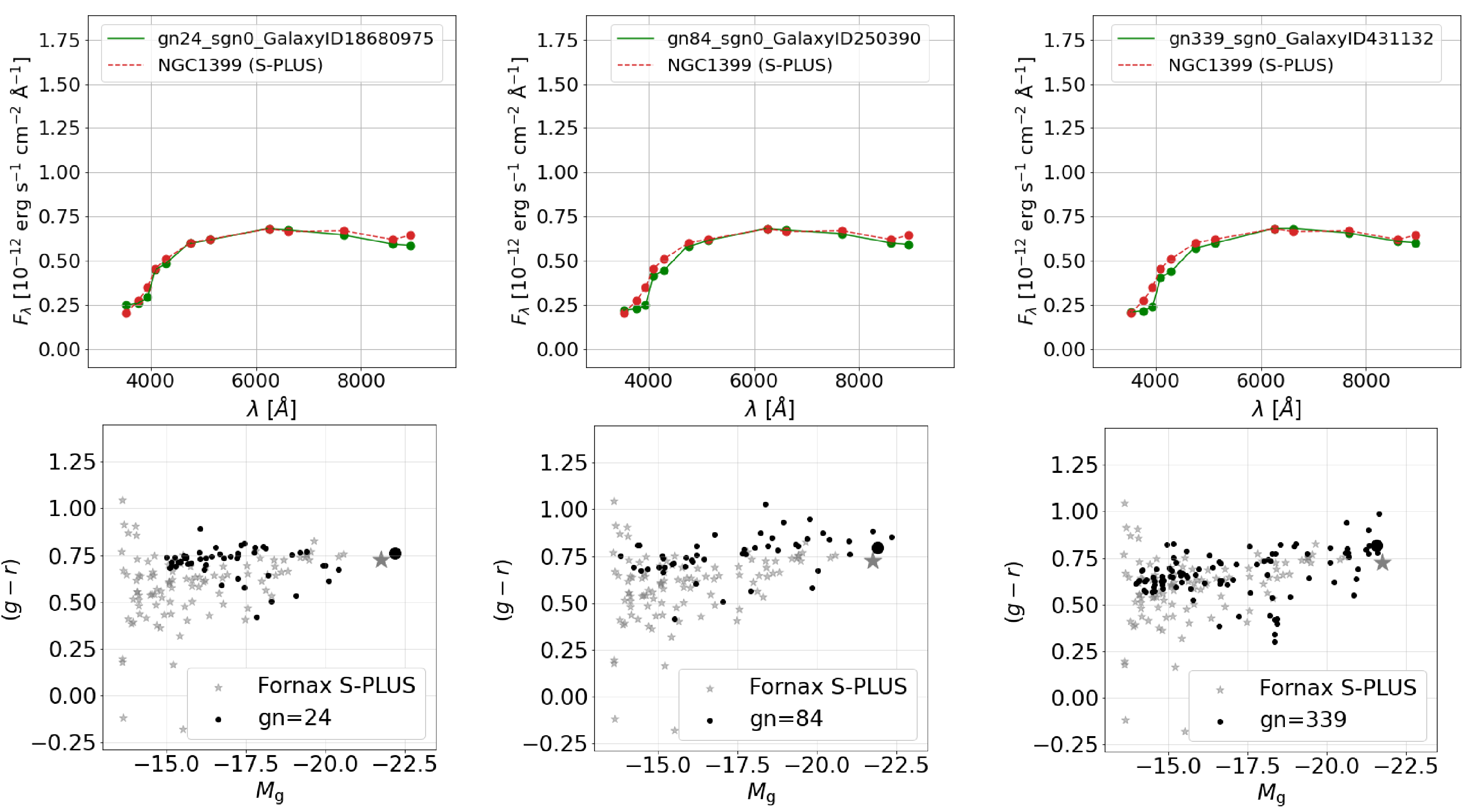}
	\caption{{\textit{Top}}: Spectral Energy Distribution (SED) of the simulated central galaxies shown in Figure\,\ref{fig:fig1_simus}. The solid green line represents the simulated photo-spectrum, while the dashed red line corresponds to the photo-spectrum of NGC\,1399 obteined with S-PLUS. {\textit{Bottom}}:  $(g-r)$ versus $M_{\rm g}$  diagram of the simulated galaxy clusters shown in Figure\,\ref{fig:fig1_simus}. For comparison, grey stars correspond to Fornax galaxies observed with S-PLUS.}
    \label{fig:fig2_simus}
\end{figure*}

\subsubsection{Simulated spectra, magnitudes and mock images}

As a first step in the analysis of Fornax-like clusters, and aiming at a comprehensive comparison between simulations and observations from S-PLUS, we used the {\sc{skirt}} radiative transfer code (\citealp{Camps2020}) to generate spectra and data cubes of every galaxy in our sample of simulated clusters. Our implementation of {\sc{skirt}} is based on the ones of \citet{Trayford2017} and \citet{RodriguezGomez2019} for {\sc{eagle}} and {\sc{IllustrisTNG}} galaxies, respectively. We remit the reader to those articles for full details, but here we provide a brief summary. For each simulated galaxy, we consider stellar sources{\footnote{Note that, in the simulations, `stellar sources' refers to star particles, with those star particles representing a simple stellar population, not an individual star.}} within a sphere of $50~\rm{kpc}$ in radius centred in its centre of potential, and divide the sources in old and young stellar populations (age $t \geqslant 100~{\rm{Myr}}$ and $t < 100~\rm{Myr}$, respectively). The SEDs of old stellar populations are modeled with GALAXEV (\citealp{Bruzual2003}), adopting a \citet{Chabrier2003} initial mass function, while SEDs of young populations are assigned MAPPINGS-III (\citealp{Groves2008}) spectral models.  

We constructed simulated spectra and datacubes using as wavelength grid the 12 S-PLUS photometric bands with their corresponding transmission curves. To obtain datacubes and mock images with the same image scale as S-PLUS ($0.55~\rm{arcsec}~\rm{pixel}^{-1}$), we use a square FoV of $16.52~\rm{arcmin}^2$, equivalent to a square area of $100~\rm{kpc}\times100~\rm{kpc}$ at the distance of Fornax, and assume $(m − M) = 31.51$ for Fornax \citep{Iodice2019}. The number of pixels of the instruments simulated by {\sc{skirt}} were adjusted accordingly, using a cubic box of 100 kpc to enclose all the simulated stellar sources. With the simulated spectra, we can compute magnitudes and colours of all the galaxies in our Fornax-like clusters, in the 12 S-PLUS bands, so those simulated quantities can be directly compared with observed data from S-PLUS. Also, using the generated datacubes, we can construct, for example, $gri$ mock images of our simulated galaxies by combining the corresponding frames of the datacubes.

In the top panels of Figure\,\ref{fig:fig1_simus}, the spatial distribution of galaxies in three of our Fornax-like clusters can be seen (one extracted from each simulation we are using). In particular, the Fornax-like cluster extracted from TNG300-1 (right panel) is the one that most resembles Fornax, as it displays two main sub-structures: the central region (resembling the central region of Fornax), and a second region outside its virial radius, resembling the surroundings of NGC\,1316 (Fornax\,A; see e.g. \citealp{SmithCastelli2021}, their figure\,1, or Figure\,\ref{pointings} and Figure\,\ref{fig.halpha-dist} in this paper). On the other hand, in the bottom panels of Figure\,\ref{fig:fig1_simus}, mock images of the central galaxy of each cluster are shown, where each image was generated by using the $gri$ S-PLUS bands and the scheme of \citet{Lupton2004}, with the same stretch and softening parameters in all galaxies.

The top panels in Figure\,\ref{fig:fig2_simus} show the comparison of the SEDs of the central galaxies in Figure\,\ref{fig:fig1_simus} with the SED of NGC 1399. The simulated spectra and the spectrum of NGC\,1399 are normalized at $\lambda = 6258$~\AA, thus becoming clear that the simulated spectra are consistent with the S-PLUS observations. The bottom panels of Figure\,~\ref{fig:fig2_simus} show the $(g-r)$ versus $M_{\rm g}$ diagram of the simulated clusters presented in Figure\,\ref{fig:fig1_simus}, compared to the CMD of the Fornax cluster. Magnitudes and colours of each simulated galaxy were estimated in the 12 S-PLUS bands using their corresponding SED. In order to improve the comparison between simulations and observations, the observed magnitudes are limited to $M_{\rm g} < -13.5$ mag. It can be seen that the simulated CMDs shown here are consistent with the CMD of Fornax obtained from S-PLUS. Although a red sequence of galaxies is clearly traced in the simulated CMDs, the blue cloud of the diagram is underpopulated, compared with the observations. Hence, the simulations predict a lower number of blue galaxies in Fornax-like clusters selected with our criteria. As we checked, the blue cloud is quite populated when selecting Fornax-like clusters according to their $M_{vir}$ and $R_{vir}$ but, when restricting the properties of the simulated central galaxies, the number density of galaxies in the blue cloud decreases. We will carry out a more in-depth analysis of this behavior in a separate paper (Zenocratti et al. in prep.).

It is worth noting that the observed colours and magnitudes were corrected for Galactic extinction and intrinsic reddening, and they were extracted from S-PLUS DR3 catalogues. S-PLUS magnitudes were measured using several apertures, optimised for different applications. Here we used apertures labelled as ‘AUTO’, defined in terms of the Kron elliptical aperture, to integrate the total flux of extended sources, being these magnitudes the most appropriate for bright objects (see \citealp{Almeida-Fernandes2022} for details).

\subsubsection{Next steps in the comparison between simulations and S-PLUS observations}

Our implementation of the {\sc{skirt}} code is currently being tested in order to obtain simulated magnitudes, colours and mock images as similar as possible to those observed with S-PLUS. With refined mock images, a robust statistical analysis of simulated spectra and magnitudes will be carried out, followed by the post-processing of simulated images (e.g. their convolution with a point spread function, and the addition of instrumental and background noises consistent with those in S-PLUS images) to estimate the structural parameters of galaxies (ellipticities, orientations, radii, etc). We aim to develop a detailed morphological analysis (both visual and quantitative), followed by a comprehensive statistical comparison between the population of galaxies in our simulated clusters with those in Fornax. Physical properties of simulated galaxies in our Fornax-like clusters (such as masses, metallicities, kinematics, stellar ages, etc) will be studied in detail and also compared to those derived from S-PLUS observations. The final step in the study of simulated Fornax-like clusters is to analyse the assembly history of our simulated clusters in order to be able to propose possible formation and evolution scenarios for the Fornax galaxy cluster.

\subsection{Synergy with other surveys and data}
\subsubsection{\text{CHANCES} of 4MOST}

The CHileAN Cluster galaxy Evolution Survey (\texttt{CHANCES}; \citealt{CHANCES}) aims at obtaining low resolution (R=5000) spectra in the 370–950 nm range for $\sim 300.000$ galaxies in 150 clusters at $0<z<0.45$   using the 4MOST facility. 4MOST is the wide-FoV (4.1 deg$^2$) fiber-fed spectrograph designed for the 4-m VISTA telescope that is foreseen to survey in 5 years, starting in 2024, most of the Southern sky, thus resulting in more than $20\times10^6$ spectra. It will be capable of obtaining, simultaneously, low-resolution spectra for $\sim$ 1,624 objects with a minimum target separation of 17 arcsec. The main goal of \texttt{CHANCES} is to study galaxy evolution in the targeted clusters out to 5 $R_{vir}$, reaching the regions were pre-processing is expected to play a key role.

\begin{figure}
    \centering
    \includegraphics[width=1.0\columnwidth]{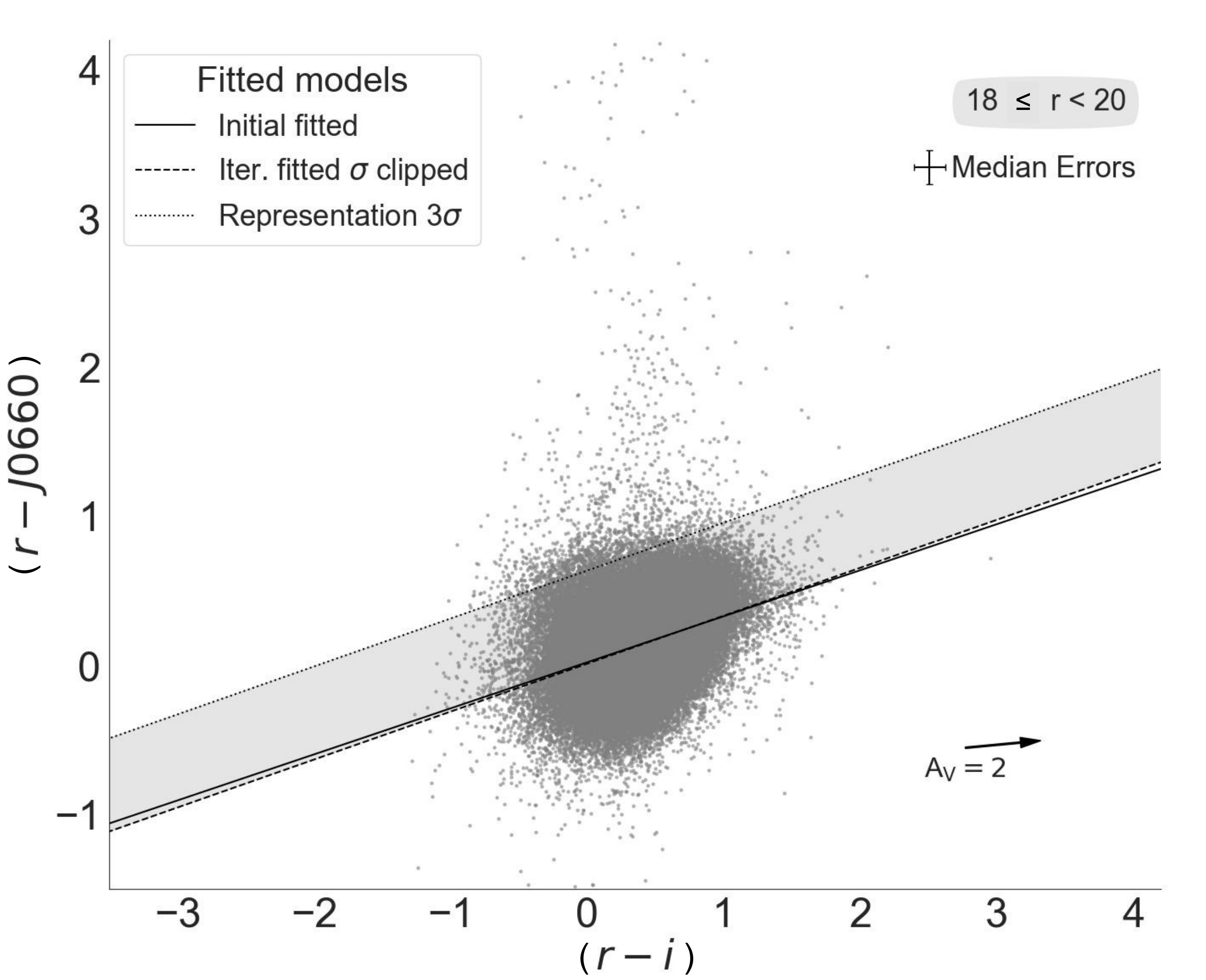}
    \caption{An illustration of the selection criteria used to identify strong
    emission-line objects using a $(r - J0660)$ versus $(r - i)$ plot. The data depicted in this plot are derived from (RUN\,1+RUN\,2)$_{res}$ in the range 18 mag $\leq r <$ 20 mag. Objects exhibiting an H$\alpha$ excess should be situated towards the upper region of this colour-colour diagram. The thin continuous line illustrates the original linear fit to all the data (grey points). The over imposed dashed line represents the final fit of the locus of points which were obtained by applying an iterative $\sigma$-clipping technique to the initial fit. The specific criterion for selecting H{$\alpha$} emitters is indicated by the upper dotted line, corresponding to a threshold of 3$\sigma$ above the final fit. Objects selected as H{$\alpha$} emitters must be above the dotted line.}
    \label{emitters-chances}
\end{figure}

\begin{figure}
    \centering
    \includegraphics[width=0.49\textwidth]{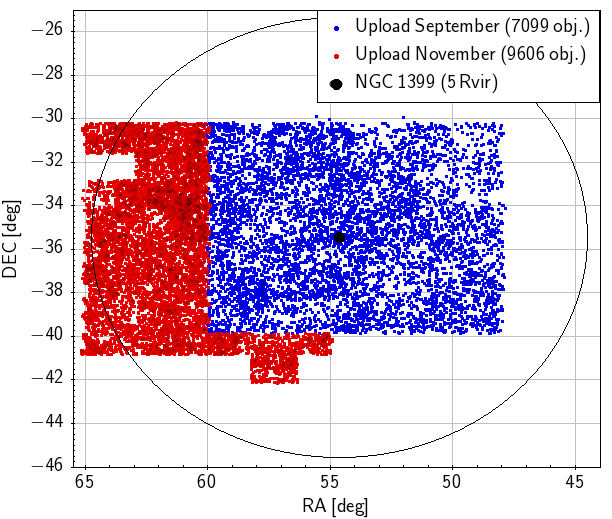}
    \caption{Progress in the sky coverage of the target selection for \texttt{CHANCES} of 4MOST. In blue, we show the region covered by the set of targets selected for the first deadline of 4MOST target upload, in September 2023. In red, we show the  coverage extension achieved in the second deadline, by the end of November 2023. We expect to reach a coverage of 5 $R_{vir}$ (big open black circle) around NGC\,1399 (black dot) for the very last deadline of 4MOST target upload, in April 2024.}
    \label{CMD-CHANCES}
\end{figure}

One of the specific clusters that \texttt{CHANCES} plans to observe is Fornax. Thanks to the generation of the specific catalogues of galaxies introduced in Section\,\ref{subsec:catalogs}, part of the S+FP team is helping to define the list of targets to be observed in Fornax within the framework of \texttt{CHANCES}. To that aim, as a first step, we cross-matched the FLS with the photometric catalogues provided by DECaLS DR10, considering a maximum matching error of 10 arcsec and taking into account that the \texttt{CHANCES} targets must be limited by r$_{LS}~<~20.5$ mag. The matching error was chosen to avoid missing extended and star-forming galaxies displaying such a difference between the reported coordinate and that obtained from the automatic photometry performed by DECaLS. From this first selection, we got a list of 196 Fornax galaxies already reported as spectroscopically confirmed members that will likely represent good targets for stellar population analysis. We also selected 273 objects without radial velocities and reported as Fornax members based on morphological criteria that, once observed by \texttt{CHANCES}, will enlarge the sample of spectroscopically confirmed members of Fornax. In addition, following \cite{Gutierrez-Soto2020}, we built the colour-colour diagram shown in Figure\,\ref{emitters-chances} with the objects included in the catalogue (RUN\,1+RUN\,2)$_{res}$. Using this plot, we selected the galaxies that detach in 3$\sigma$ from the grey zone which are expected to be H$\alpha$ emitters. After cleaning this selection from spectroscopically confirmed background objects, we got a small sample of 14 objects without radial velocity in the literature and displaying distorted morphologies that could be associated to objects falling into the cluster or experiencing stripping processes. As a final step, we restricted a bit more the (RUN\,1+RUN\,2)$_{res}$ catalogues using additional constraints like restricting \texttt{SExtractor} flag values and broad-bands magnitudes errors, and we also cleaned the catalog from spectroscopically confirmed background galaxies. In that way, we obtained an additional sample of $\sim$16,700 objects covering the area shown in Figure\,\ref{CMD-CHANCES}. At the moment we are still working on extending the last selection of galaxies in order to reach 5 R$_{vir}$ around NGC\,1399.

\subsubsection{The Fornax3D survey and MUSE data}

Beside the aforementioned checks on the S-PLUS H$\alpha$+[NII] flux estimates (Section\,\ref{subsec:elgs}), the MUSE data from the Fornax3D survey (and other Fornax targets) can be combined with the S-PLUS photometric measurements to also deliver a better picture for the star-formation history (SFH) of Fornax cluster galaxies and the role of such an environment in driving it. Indeed, although the 4750 - 9300 \AA~MUSE wavelength range includes a considerable range of stellar absorption features, it still misses some critical stellar population age indicators at the blue end of the optical spectrum such as the D4000 Balmer break and H$\delta$ (e.g., \citealt{Kauffmann2003}), which are critical for constraining the role of the most recent star-formation events. These limitations were discussed in \citet{Loni2023} for the particularly interesting case of the Fornax spiral galaxy NGC\.1436, which was stripped of its extended HI reservoir after entering the cluster while still showing central star formation. Indeed, although their study allowed to uncover how this object experienced an intense burst of star formation as it first entered the cluster, the MUSE data do not allow to firmly constrain the most recent phases of such a transformation process since they make the SFH analysis over-reliant on Balmer absorption lines, which do not have a monotonic dependence on age for young and intermediate-age stellar populations. The advent of spectro-photometric fitting techniques (e.g., \citealt{Lopez-Fernandez2016,Cappellari2023}) will allow to combine our S-PLUS measurements with MUSE data for the Fornax galaxies and overcome these limitations, tapping in particular on the bluer S-PLUS bands that cover the Balmer break. Obtaining in this way a complete picture for the SFH of Fornax galaxies is particularly compelling considering how deep VST optical and MeerKAT radio data (from \citealt{Iodice2019,Serra2023}, respectively) will allow to place these results in relation to on-going evidence of gravitational and hydrodynamical perturbations, thus adding to our understanding of how such processes drive the morphological transformation of galaxies in intermediate-mass clusters such as Fornax.

\section{Concluding Remarks}
\label{sec:conclusions}

The Fornax cluster is a remarkable laboratory to study in detail a variety of physical processes that contribute to shaping the observed properties of galaxies in high-density environments. Its close distance makes it possible to observe in great detail structural components and spectrophotometric properties of cluster galaxies down to very low levels of surface brightness, including the GC systems associated with them. The dynamically young nature of Fornax turns out to be of great advantage to try to effectively probe a variety of processes. More specifically, the known substructures, their dynamics, and the overall PPS distribution of the different galaxy populations in the cluster offer the possibility of identifying under what conditions and timescales one specific process may dominate over another. This is still an outstanding problem in environmental evolution studies of galaxies.  

The S+FP has been especially designed to take full advantage of the extensive photometric information delivered by the 12-filter system of S-PLUS. We have performed a number of sanity checks, including: the recovery of  $M_{vir}$ and $R_{vir}$ values for Fornax, the resulting S-PLUS broad-band integrated magnitudes for galaxies with 10.5 mag $\lesssim g \lesssim$ 20.5 mag in consonance with those obtained by deeper surveys (e.g., DECaLS) and a Fornax luminosity function steeper than that in the field at the low-luminosity end in consonance with other works in the literature.  In this paper we present different aspects of the science that can be undertaken with this dataset, including the following:

(1) The combination of narrow- and broad-band filters in S-PLUS  offers, among other possibilities, the identification of line emitters that can be used as tracers of the star-forming activity in the cluster. This is particularly relevant to put together a consistent picture of how the higher density environment of a system like Fornax can impact the star formation history of galaxies, ultimately leading to the star formation quenching in them. Our analysis shows that although H$\alpha$ emitters are more present in low-density regions of the cluster, there is a clear bimodal distribution, with the denser population of H$\alpha$ emitters showing a higher fraction of early-type galaxies and the less dense population presenting a higher fraction of late type and peculiars and irregular morphologies.

(2) S-PLUS allows us to reach low levels of surface brightness, opening the possibility for the S+FP to detect LSB galaxies and UDGs in Fornax. Together with the already-known samples of LSB galaxies and UDGs in other clusters such as Coma \citep{vanDokkum_2015}, the existence of this type of galaxies in high-density environments raise fundamental questions regarding their origin and later evolution. Their associated GC systems have been proposed as tools to cast some light on those issues. We demonstrate the ability of the S+FP to detect GCs in Fornax, opening up the way for a better understanding of the formation and evolution of LSB galaxies and UDGs.

(3) AI techniques together with state-of-the-art simulations are being used by the S+FP to further the scientific exploitation and interpretation of the massive amount of photometric data delivered by S-PLUS on Fornax. While AI provides more efficient and objectives ways of analysing and classifying the available information, detailed simulations allow us to further understand the physics behind the observed galaxy properties. This is a fundamental aspect of the full characterization of Fornax, as a whole, and of the different cluster galaxy populations. We identify  a set of 45 Fornax-like clusters within the {\sc eagle} and {\sc IllustrisTNG} TNG100, TNG300 cosmological simulations and create mock images on all 12 S-PLUS bands of these structures down to galaxy stellar masses of $M_\star \geqslant 10^8~{\rm{M}}_\odot$. Our findings show that the stellar populations for the central galaxies in these simulated structures — based on colour-magnitude diagrams and SEDs —  reproduce S+FP analysis for Fornax’ central galaxy NGC\,1399. 

The above set of results is part of an on-going effort that is constantly in the process of improving different analysis and interpretation tools. Even in this first stage, the S+FP is generating crucial material to seed fruitful collaborative efforts, as it was shown in the case of \texttt{CHANCES} of 4MOST, but, also, for the whole astronomical community. This paper presents a number of preliminary works that merely represent the tip of the iceberg of what can be done. We plan to produce more advanced results that will be part of a series of papers currently in preparation deepening the different topics addressed in Section\,\ref{sec:results}.

\section*{Acknowledgements}
We would like the thank the anonymous referee for her/his detailed revision and useful analysis of this paper that greatly helped to improve its content.
AVSC, RFH, ARL, JPC, LZ, FRF and LAGS acknowledge financial support from CONICET, Agencia I+D+i (PICT 2019-03299) and Universidad Nacional de La Plata (Argentina). AC acknowledge the financial support provided by FAPERJ grants E26/200.607/2002 e 210.371/2022(270993). LLN thanks Funda\c{c}\~ao de Amparo \`a Pesquisa do Estado do Rio de Janeiro (FAPERJ) for granting the postdoctoral research fellowship E-40/2021(280692). 
LSJ acknowledges the support from CNPq (308994/2021-3) and FAPESP (2011/51680-6). MEDR acknowledges support from PICT-2021-GRF-TI-00290 of Agencia I+D+i (Argentina). FRH thanks FAPESP's support with the grant 2018/21661-9. RD gratefully acknowledges support by the ANID BASAL project FB210003. NWCL gratefully acknowledges the generous support of a Fondecyt General grant 1230082, as well as support from Núcleo Milenio NCN2023\_002 (TITANs) and funding via the BASAL Centro de Excelencia en Astrofisica y Tecnologias Afines (CATA) grant PFB-06/2007.  NWCL also thanks support from ANID BASAL project ACE210002 and ANID BASAL projects ACE210002 and FB210003. YJT acknowledges financial support from the European Union’s Horizon 2020 research and innovation programme under the Marie Skłodowska-Curie grant agreement No 898633, the MSCA IF Extensions Program of the Spanish National Research Council (CSIC), the State Agency for Research of the Spanish MCIU through the Center of Excellence Severo Ochoa award to the Instituto de Astrofísica de Andalucía (SEV-2017-0709), and grant CEX2021-001131-S funded by MCIN/AEI/ 10.13039/501100011033. MG acknowledges support from FAPERJ grant E-26/211.370/2021. JAHJ acknowledges support from FAPESP, process number 2021/08920-8. ACK thanks FAPESP for the support grant 2020/16416-5 and to CNPq. DEOR acknowledges the financial support from the Chilean National Agency for Research and Development (ANID), InES-G\'{e}nero project INGE210025. CMdO thanks support from FAPESP grant 2019/26492-3 and CNPq grant number 309209/2019-6. 

The S-PLUS project, including the T80-South robotic telescope and the S-PLUS scientific survey, was founded as a partnership between the Fundação de Amparo à Pesquisa do Estado de São Paulo (FAPESP), the Observatório Nacional (ON), the Federal University of Sergipe (UFS), and the Federal University of Santa Catarina (UFSC), with important financial and practical contributions from other collaborating institutes in Brazil, Chile (Universidad de La Serena), and Spain (Centro de Estudios de Física del Cosmos de Aragón, CEFCA). We further acknowledge financial support from the São Paulo Research Foundation (FAPESP), Fundação de Amparo à Pesquisa do Estado do RS (FAPERGS), the Brazilian National Research Council (CNPq), the Coordination for the Improvement of Higher Education Personnel (CAPES), the Carlos Chagas Filho Rio de Janeiro State Research Foundation (FAPERJ), and the Brazilian Innovation Agency (FINEP). The authors who are members of the S-PLUS collaboration are grateful for the contributions from CTIO staff in helping in the construction, commissioning and maintenance of the T80-South telescope and camera. We are also indebted to Rene Laporte and INPE, as well as Keith Taylor, for their important contributions to the project. From CEFCA, we particularly would like to thank Antonio Marín-Franch for his invaluable contributions in the early phases of the project, David Cristóbal-Hornillos and his team for their help with the installation of the data reduction package jype version 0.9.9, César Íñiguez for providing 2D measurements of the filter transmissions, and all other staff members for their support with various aspects of the project.

The Legacy Surveys consist of three individual and complementary projects: the Dark Energy Camera Legacy Survey (DECaLS; Proposal ID 2014B-0404; PIs: David Schlegel and Arjun Dey), the Beijing-Arizona Sky Survey (BASS; NOAO Prop. ID 2015A-0801; PIs: Zhou Xu and Xiaohui Fan), and the Mayall $z$-band Legacy Survey (MzLS; Prop. ID 2016A-0453; PI: Arjun Dey). DECaLS, BASS and MzLS together include data obtained, respectively, at the Blanco telescope, Cerro Tololo Inter-American Observatory, NSF’s NOIRLab; the Bok telescope, Steward Observatory, University of Arizona; and the Mayall telescope, Kitt Peak National Observatory, NOIRLab. Pipeline processing and analyses of the data were supported by NOIRLab and the Lawrence Berkeley National Laboratory (LBNL). The Legacy Surveys project is honored to be permitted to conduct astronomical research on Iolkam Du’ag (Kitt Peak), a mountain with particular significance to the Tohono O’odham Nation.

NOIRLab is operated by the Association of Universities for Research in Astronomy (AURA) under a cooperative agreement with the National Science Foundation. LBNL is managed by the Regents of the University of California under contract to the U.S. Department of Energy.

This project used data obtained with the Dark Energy Camera (DECam), which was constructed by the Dark Energy Survey (DES) collaboration. Funding for the DES Projects has been provided by the U.S. Department of Energy, the U.S. National Science Foundation, the Ministry of Science and Education of Spain, the Science and Technology Facilities Council of the United Kingdom, the Higher Education Funding Council for England, the National Center for Supercomputing Applications at the University of Illinois at Urbana-Champaign, the Kavli Institute of Cosmological Physics at the University of Chicago, Center for Cosmology and Astro-Particle Physics at the Ohio State University, the Mitchell Institute for Fundamental Physics and Astronomy at Texas A\&M University, Financiadora de Estudos e Projetos, Fundacao Carlos Chagas Filho de Amparo, Financiadora de Estudos e Projetos, Fundacao Carlos Chagas Filho de Amparo a Pesquisa do Estado do Rio de Janeiro, Conselho Nacional de Desenvolvimento Cientifico e Tecnologico and the Ministerio da Ciencia, Tecnologia e Inovacao, the Deutsche Forschungsgemeinschaft and the Collaborating Institutions in the Dark Energy Survey. The Collaborating Institutions are Argonne National Laboratory, the University of California at Santa Cruz, the University of Cambridge, Centro de Investigaciones Energeticas, Medioambientales y Tecnologicas-Madrid, the University of Chicago, University College London, the DES-Brazil Consortium, the University of Edinburgh, the Eidgenossische Technische Hochschule (ETH) Zurich, Fermi National Accelerator Laboratory, the University of Illinois at Urbana-Champaign, the Institut de Ciencies de l’Espai (IEEC/CSIC), the Institut de Fisica d’Altes Energies, Lawrence Berkeley National Laboratory, the Ludwig Maximilians Universitat Munchen and the associated Excellence Cluster Universe, the University of Michigan, NSF’s NOIRLab, the University of Nottingham, the Ohio State University, the University of Pennsylvania, the University of Portsmouth, SLAC National Accelerator Laboratory, Stanford University, the University of Sussex, and Texas A\&M University.

BASS is a key project of the Telescope Access Program (TAP), which has been funded by the National Astronomical Observatories of China, the Chinese Academy of Sciences (the Strategic Priority Research Program “The Emergence of Cosmological Structures” Grant XDB09000000), and the Special Fund for Astronomy from the Ministry of Finance. The BASS is also supported by the External Cooperation Program of Chinese Academy of Sciences (Grant  114A11KYSB20160057), and Chinese National Natural Science Foundation (Grant  12120101003,  11433005).

The Legacy Survey team makes use of data products from the Near-Earth Object Wide-field Infrared Survey Explorer (NEOWISE), which is a project of the Jet Propulsion Laboratory/California Institute of Technology. NEOWISE is funded by the National Aeronautics and Space Administration.

The Legacy Surveys imaging of the DESI footprint is supported by the Director, Office of Science, Office of High Energy Physics of the U.S. Department of Energy under Contract No. DE-AC02-05CH1123, by the National Energy Research Scientific Computing Center, a DOE Office of Science User Facility under the same contract; and by the U.S. National Science Foundation, Division of Astronomical Sciences under Contract No. AST-0950945 to NOAO.

This research has made use of NASA’s Astrophysics Data System (ADS) Bibliographic Services. The ADS is operated by the Smithsonian Astrophysical Observatory under NASA Cooperative Agreement 80NSSC21M0056.

This research has made use of the NASA/IPAC Extragalactic Database (NED), which is funded by the National Aeronautics and Space Administration and operated by the California Institute of Technology.

This research has made use of \texttt{TOPCAT} \citep{TOPCAT}. TOPCAT was initially (2003-2005) developed under the UK Starlink project (1980-2005, R.I.P.). Since then it has been supported by grant PP/D002486/1 from the UK's Particle Physics and Astronomy Research Council, the VOTech project (from EU FP6), the AstroGrid project (from PPARC/STFC), the AIDA project (from EU FP7), grants ST/H008470/1, ST/I00176X/1, ST/J001414/1 ST/L002388/1, ST/M000907/1, ST/R000700/1, ST/S001980/1, ST/V000454/1 and ST/X002969/1 from the UK's Science and Technology Facilities Council (STFC), the GAVO project (BMBF Bewilligungsnummer 05A08VHA), the European Space Agency, the FP7 project GENIUS, and Europlanet-2024 RI (EU Horizon 2020 grant agreement No 871149).

\section*{Data Availability}

The data underlying this article will be shared on reasonable request to the corresponding author.



\bibliographystyle{mnras}
\bibliography{S+FP_I} 








\bsp	
\label{lastpage}
\end{document}